\documentclass[preprint,12pt]{elsarticle}

\usepackage{amssymb}
\usepackage[utf8]{inputenc} 
\usepackage[T1]{fontenc}    
\usepackage{url}            
\usepackage{booktabs}       
\usepackage{amsfonts}       
\usepackage{nicefrac}       
\usepackage{microtype}      
\usepackage{lipsum}
\usepackage{fancyhdr}       
\usepackage{graphicx}       
\graphicspath{{media/}}     
\usepackage{caption}
\usepackage{subcaption}
\usepackage{diagbox}
\usepackage{threeparttable}
\usepackage[linesnumbered,ruled,vlined]{algorithm2e}
\usepackage{multirow}
\usepackage{ulem}
\usepackage{amsmath}
\usepackage{float}
\usepackage{natbib}
\usepackage{mathtools}
\mathtoolsset{showonlyrefs=False}
\usepackage[acronym]{glossaries}
\usepackage{acronym}
\usepackage[table,xcdraw]{xcolor}
\usepackage{todonotes}
\usepackage{adjustbox}

\usepackage{siunitx}
\usepackage{makecell}
\usepackage{multirow, multicol}
\renewrobustcmd{\bfseries}{\fontseries{b}\selectfont}
\renewrobustcmd{\boldmath}{}
\newrobustcmd{\B}{\bfseries}

\journal{Information Sciences}

\makeglossaries
\newacronym{nir}{NIR}{neural information retrieval}
\newacronym{plm}{PLM}{pretrained language model}
\acrodefplural{plm}[PLMs]{pretrained language models}
\newacronym{drmm}{DRMM}{Deep Relevance Matching Model}
\newacronym{knrm}{KNRM}{Kernel-based Neural Ranking Model}
\newacronym{bertdot}{BERTdot}{BERT with Dot Production}
\newacronym{colbert}{ColBERT}{Contextualized Late Interaction over BERT}
\newacronym{ewc}{EWC}{Elastic Weight Consolidation}
\newacronym{ewcol}{EWCol}{Elastic Weight Consolidation in the Online Paradigm}
\newacronym{si}{SI}{Synaptic Intelligence}
\newacronym{mas}{MAS}{Memory Aware Synapses}
\newacronym{nr}{NR}{Naive Rehearsal}
\newacronym{gem}{GEM}{Gradient Episodic Memory}
\newacronym{clnir}{CLNIR}{Continual Learning Framework for Neural Information Retrieval}
\newacronym{mrr}{MRR}{mean reciprocal rank}
\newacronym{fwt}{FWT}{forward transfer}
\newacronym{bwt}{BWT}{backward transfer}

\newglossaryentry{l2}
{
    name=L2,
    description={L2 regularization}
}
\newglossaryentry{duet}
{
    name=DUET,
    description={Duet architecture}
}

\usepackage{url}
\usepackage{hyperref}%
\hypersetup{%
  breaklinks=true,%
  linktocpage=true,
  colorlinks=true,%
  linkcolor=blue,%
  citecolor=blue,%
  filecolor=blue,%
  urlcolor=blue,%
  pdftitle={},%
  pdfsubject={},%
  pdfauthor={},%
  pdfkeywords={},%
  pdfproducer={}%
  }%

\newcommand{\singleDocument}{d}

\newcommand{\singleQuery}{q}
\newcommand{\singleQueryTokens}{\mathbf{w}_\singleQuery}
\newcommand{\singleDocumentTokens}{\mathbf{w}_\singleDocument}
\newcommand{\singleQueryMask}{\mathbf{m}_\singleQuery}
\newcommand{\singleDocumentMask}{\mathbf{m}_\singleDocument}
\newcommand{\singleQueryEmbeddings}{\mathbf{E}_\singleQuery}
\newcommand{\singleDocumentEmbeddings}{\mathbf{E}_\singleDocument}
\newcommand{\singleQueryBertEmbeddings}{\mathbf{B}_\singleQuery}
\newcommand{\singleDocumentBertEmbeddings}{\mathbf{B}_\singleDocument}

\newcommand{\queryBatch}{\mathbf{q}}
\newcommand{\documentBatchPositive}{\mathbf{pos}}
\newcommand{\documentBatchNegative}{\mathbf{neg}}
\newcommand{\margin}{\mathit{margin}}

\newcommand{\documentLength}{\mathit{l}(\singleDocument)}
\newcommand{\queryLength}{\mathit{l}(\singleQuery)}
\newcommand{\embeddingDimension}{n}

\newcommand{\pairwiseRankingSample}{\langle \queryBatch, \documentBatchPositive, \documentBatchNegative \rangle}

\newcommand{\taskIndex}{t}

\newcommand{\nTasks}{T}

\newcommand{\lossFunction}{\ell}
\newcommand{\ranker}{\mathcal{R}}
\newcommand{\learningStrategy}{\mathcal{L}}

\newcommand{\performance}{P}
\newcommand{\finalPerformance}{\performance_{\mathrm{final}}}

\newcommand{\bwt}{\mathit{BWT}}
\newcommand{\fwt}{\mathit{FWT}}

\newcommand{\reals}{\mathbb{R}}
\newcommand{\agentModel}{\mathcal{A}}
\newcommand{\regularizationFunction}{\Omega}
\newcommand{\memorySet}{\mathbf{M}}

\newcommand{\regularizationStrategySet}{\mathit{regularization}}
\newcommand{\replayStrategySet}{\mathit{replay}}
\newcommand*{\tran}{^{\mkern-1.5mu\mathsf{T}}}

\begin{document}

\begin{frontmatter}


\title{Advancing continual lifelong learning in neural information retrieval: definition, dataset, framework, and empirical evaluation}


\author[aff1]{Jingrui Hou}
\ead{j.hou@lboro.ac.uk}
\author[aff1]{Georgina Cosma\corref{cor1}}
\ead{g.cosma@lboro.ac.uk}
\author[aff2]{Axel Finke}
\ead{a.finke@lboro.ac.uk}
\cortext[cor1]{Corresponding author}
\affiliation[aff1]{organization={Department of Computer Science, School of Science, Loughborough University},
            addressline={Epinal Way}, 
            city={Loughborough},
            postcode={LE11 3TU}, 
            state={Leicestershire},
            country={UK}}
\affiliation[aff2]{organization={Department of Mathematical Sciences, School of Science, Loughborough University},
            addressline={Epinal Way}, 
            city={Loughborough},
            postcode={LE11 3TU}, 
            state={Leicestershire},
            country={UK}}

\begin{abstract}
    Continual learning refers to the capability of a machine learning model to learn and adapt to new information, without compromising its performance on previously learned tasks. Although several studies have investigated continual learning methods for \gls{nir} tasks, a well-defined task definition is still lacking, and it is unclear how typical learning strategies perform in this context. To address this challenge, a systematic task definition of continual \gls{nir} is presented, along with a multiple-topic dataset that simulates continuous information retrieval. A comprehensive continual neural information retrieval framework consisting of typical retrieval models and continual learning strategies is then proposed. Empirical evaluations illustrate that the proposed framework can successfully prevent catastrophic forgetting in neural information retrieval and enhance performance on previously learned tasks. The results also indicate that embedding-based retrieval models experience a decline in their continual learning performance as the topic shift distance and dataset volume of new tasks increase. In contrast, pretraining-based models do not show any such correlation. Adopting suitable learning strategies can mitigate the effects of topic shift and data augmentation in continual neural information retrieval.
\end{abstract}



\end{frontmatter}

\glsresetall
\section{Introduction} 

Information retrieval, a fundamental area of research within natural language processing, focuses on identifying and extracting relevant information from a collection of documents~\cite{Ceri2013}. Information retrieval technologies have led to the development of various real-world applications, including search engines, question-answering systems, and e-commerce recommender platforms~\cite{guo2020deep}. With the advancement of deep learning, neural network-based information retrieval approaches known as \textit{\gls{nir}}~\cite{Ceri2013} have exhibited superior results. Earlier \gls{nir} models were primarily based on word embedding methods, while more recent \gls{nir} models~\cite{zhao2024dense} have utilised \textit{\glspl{plm}}~\cite{DBLP:conf/naacl/DevlinCLT19} for improved performance. 

Most deep learning models are trained using a fixed dataset, known as the \textit{isolated learning paradigm}~\cite{liu2017lifelong}, and \gls{nir} is no exception. However, in real-world information retrieval applications and systems, new information is constantly arriving. Taking a document retrieval system as an example, the internal neural ranking model trained on old information may struggle to accurately sort and recommend the latest content, as new articles are constantly injected into the system over time, potentially leading to suboptimal user experiences due to the presence of different topics in the new articles. One common approach to addressing this issue is by retraining a new model each time novel data emerge. However, this practice can be considered impractical due to its low efficiency and high resource consumption~\cite{harun2023efficient}. As a consequence, the \textit{continual learning paradigm} strives to facilitate the seamless integration of new data streams into neural networks as they become available~\cite{pratama2019deep,biesialska2020continual,10444954}. The main challenge in continual learning is to avoid \textit{catastrophic forgetting}, i.e.\ the problem that deep learning models tend to inadvertently forget previously acquired knowledge when learning on new data~\cite{french1999catastrophic}.

Most existing continual learning strategies are designed for classification tasks~\cite{de2021continual}. There has been limited focus on continual learning in information retrieval~\cite{10.1007/978-3-030-72113-8_25, gerald2022continual,10.1145/3583780.3614821}, and it is not yet clear how existing learning strategies perform in this field of information retrieval. Hence, the primary objective of this study is to bridge this research gap. Moreover, the performance of continual learning can be affected by two important factors: data volume and topic shift. Data volume, i.e.\ the size of the new dataset, can lead to varying degrees of catastrophic forgetting as the amount of data increases~\cite{karakida2021learning}. On the other hand, topic shift, i.e.\ dissimilarity between new tasks and previously learned tasks, can also impact continual learning performance~\cite{pmlr-v139-lee21e}.  However, the effect of these two factors on continual learning in information retrieval tasks has not been investigated. Therefore, this study aims to fill this gap.

Considering the underutilization of continual learning strategies in information retrieval tasks and a lack of understanding regarding the impact of data volume and topic shift on continual learning performance within this domain, this study proposes multiple research objectives. The primary objective is to design a framework that incorporates typical continual learning strategies into multiple \gls{nir} methods.  The second objective involves determining the most effective continual learning strategies tailored to different \gls{nir} models. Lastly, this study investigates how variations in data volume and topic shift influence continual \gls{nir} performance.

However, it is important to note that, as of the present time, a well-defined mathematical task definition and a benchmark dataset for \gls{nir} in a continual learning context are still absent. This absence poses a challenge to achieving the aforementioned research objectives effectively. In this paper, significant groundwork is laid out, which includes a precise task definition, the establishment of a benchmark dataset, and the development of a comprehensive continual \gls{nir} framework. These foundational elements serve as prerequisites for the empirical evaluation of the proposed research objectives. In summary, the key contributions of this paper are as follows:
\begin{enumerate}
    \item A precise definition of the continual learning paradigm in the context of \gls{nir} tasks, addressing the lack of a standardized definition tailored to continual learning in this domain.
    
    \item  A new dataset, called \textit{Topic-MSMARCO}, specifically designed for evaluating continual \gls{nir} tasks. {Topic-MSMARCO includes multiple thematic \gls{nir} tasks and predefined task similarity, filling the gap in the availability of evaluation benchmarks tailored to continual learning in \gls{nir}.
    
    \item  A novel framework named \textit{\gls{clnir}}. By allowing for the pairing of different models and strategies, \gls{clnir} offers a flexible and customizable approach to continual \gls{nir}, generating a wide array of continual learning methods. This framework provides researchers with a comprehensive baseline for further exploration.}
    
    \item An empirical evaluation of the performance of different continual learning strategies for various \gls{nir} models, as well as an investigation into the impact of topic shift distance and data volume on continual \gls{nir} performance.
\end{enumerate}

The remainder of the paper is organized as follows: Section~\ref{section2} provides an overview of existing continual learning strategies and their attempted application in the field of \gls{nir}, thus laying the foundations for the proposed research. Section~\ref{section3} outlines the task definition and the evaluation metrics used in this study. Section~\ref{section4} presents the proposed \gls{clnir} and its architecture. Section~\ref{section5} details the proposed {\it Topic-MSMARCO} dataset and the experiments on evaluating the two research objectives. Lastly, Section~\ref{section6} concludes this study and provides suggestions for future work.

\section{Related work}\label{section2}

This section provides an overview of existing research on continual learning strategies and their application to information retrieval tasks.

\subsection{Continual learning strategies}\label{litcon1}
Continual learning can be classified into three categories: \textit{regularization-based}, \textit{replay-based}, and \textit{parameter isolation} methods~\cite{de2021continual}.

\textbf{Regularization-based strategies} primarily employ penalty terms in loss functions to control parameter updates. A well-known example of this approach is \textit{\gls{ewc}} \cite{kirkpatrick2017overcoming}, which slows down learning on certain neural parameter weights based on their importance to previous tasks. \citet{8545895} enhanced \gls{ewc} with a reparameterization method utilizing the Fisher information matrix of the network parameters. Another algorithm in this category is \textit{\gls{si}} \cite{zenke2017continual}, which computes per-synapse consolidation strength in an online fashion over the entire learning trajectory in parameter space. Moreover, \citet{chaudhry2018riemannian} further combined \gls{ewc} and \gls{si}, and the fused method demonstrated superior continual learning performance. To prevent important knowledge related to previous tasks from being overwritten, the \textit{\gls{mas}} strategy \cite{aljundi2018memory} stores important information in a separate memory during training on new tasks and uses a regularization term to encourage the model to learn new information without disrupting the important information stored in the memory module. Additionally, \citet{mazur2022target} introduced a regularization strategy utilizing the Cramer-Wold distance between target layer distributions to represent both current and past information. This approach effectively addresses the problem of increasing memory consumption commonly encountered in similar methods. In order to diminish the upper bound of forgetting in regularization-based methods, \citet{10190202} investigated the geometric properties of the approximation loss of prior tasks concerning the maximum eigenvalue. For pretrained NLP tasks, \citet{zhang2023lifelong} introduced the lifelong language method with adaptive uncertainty regularization, enabling a single BERT model to dynamically adapt to various tasks without performance degradation.

\textbf{Replay-based approaches} typically store samples from old tasks and retrain them in new tasks. One example of this approach is the  \textit{incremental classifier and representation learning} \citep{rebuffi2017icarl}, which uses a herding-based step for prioritized exemplar selection and only requires a small number of exemplars per class. Another replay-based method is \textit{\gls{gem}} \citep{lopez2017gradient}, which uses a subset of observed examples to minimize negative backward transfer. \citet{ZHUANG2022108907} have proposed a continual learning technique that selects samples by combining various criteria such as distance to prototype, intra-class cluster variation, and classifier loss. Another type of replay is \textit{experience replay}. One typical method is \textit{continual learning with experience and replay (CLEAR)} \citep{rolnick2019experience}, which employs on-policy learning on fresh experiences to adapt rapidly to new tasks while utilizing off-policy learning with behavioural cloning on replay experience to maintain and modestly enhance performance on past tasks. \citet{li2024adaer} further proposed the \textit{adaptive experience replay (AdaER)}, which enhances existing experience replay by introducing a novel contextually-cued recall strategy for the replay stage and entropy-balanced reservoir sampling for the update stage.

\textbf{Parameter isolation methods} allocate dynamic network parameters to incremental data or tasks. One example of this approach is the \textit{progressive neural network} \citep{rusu2016progressive}, which prevents catastrophic forgetting by instantiating a new neural network for each task being solved. Another parameter isolation method is the \textit{expert gate} \citep{aljundi2017expert} which addresses continual learning issues by adding a new expert model when a new task arrives or knowledge is transferred from previous models. Furthermore, in dynamic graph learning, the parameter isolation graph neural network, introduced by \citet{10.1145/3539618.3591652}, is designed for continual learning on dynamic graphs, overcoming tradeoffs through parameter isolation and expansion.

\subsection{Attempts of continual learning in information retrieval}
\label{litcon2}

A limited number of studies have explored the application of continual learning in information retrieval tasks. \citet{wang2021continual} and \citet{SONG2023109276} demonstrated the impact of continual learning methods on multi-modal embedding spaces in cross-modal retrieval tasks. In the text modality, \citet{10.1007/978-3-030-72113-8_25} investigated the catastrophic forgetting problem in neural ranking models for information retrieval and their experiments revealed that the effectiveness of neural ranking comes at the cost of forgetting. However, transformer-based models provided a good balance between effectiveness and memory retention; and \gls{ewc} \citep{kirkpatrick2017overcoming} effectively mitigated catastrophic forgetting while maintaining a good trade-off between multiple tasks. \citet{gerald2022continual} built a sequential information retrieval dataset using the \textit{MSMARCO} dataset \citep{nguyen2016ms} with sentence embedding and evaluated two pretraining-based \gls{nir} models: \textit{Vanilla BERT} \citep{DBLP:conf/naacl/DevlinCLT19} and \textit{MonoT5} \citep{nogueira-etal-2020-document}. \citet{gerald2022continual} also found that catastrophic forgetting does exist in information retrieval, but to a lesser extent compared to some classification tasks. Furthermore, in the general retrieval scenario, \citet{10.1145/3583780.3614821} proposed the \textit{continual-learner for generative retrieval (CLEVER)}. CLEVER introduces Incremental Product Quantization for encoding new documents and a memory-augmented learning mechanism to retain previous knowledge. However, this method only considers the emergence of new documents.

Several studies have explored continual learning in \gls{nir} from different perspectives. Nevertheless, a precise definition of continual \gls{nir} and strategies compatible with multiple \gls{nir} methods is still lacking. This study aims to fill this research gap.

\section{Continual \gls{nir} task definition and evaluation metrics}
\label{section3}

This section introduces the task definition of continual learning in \gls{nir}, followed by a description of the metrics used to evaluate the task.

\subsection{Task definition}\label{section3.1}

A continual neural information retrieval task is a specific type of continual learning task that involves a model that is able to retrieve relevant information from a continuously growing set of queries and documents. 
\begin{itemize}

   \item Let $T \in \mathbb{N}$ be the total number of tasks.
  
    \item For each task $t \in \{1,\dotsc, T\}$, let $Q_t$ be a set of queries and let $D_t$ be a set of documents.
    \item For each task $t \in \{1,\dotsc, T\}$, let $S_t$ be a (training) dataset for \gls{nir}, i.e., for some finite number of samples $n_t$, $S_t = \{(x_t^i, y_t^i)\}_{i=1}^{n_t}$, where
    \begin{itemize}
      \item $x_t^i \coloneqq (q_t^i, d_t^i) \in Q_t \times D_t$ is the $i$th query--document pair for the $t$th task;
      \item $y_t^i \in \mathbb{R}$ is the ground-truth relevance score associated with this query--document pair.
    \end{itemize}
    Relevance scores can be either a discrete value (``relevant'' or ``irrelevant'') or a continuous value (degree of relevance, e.g.\ between 0 and 1). 
   
   \item A (training) dataset for continual \gls{nir} is then given by $\mathcal{S} \coloneqq \{S_1, \dotsc, S_\nTasks\}$.

\end{itemize}
The continual neural information retrieval task requires parameters $\theta$ of a neural information retrieval model $\ranker(\theta)$ to be learned from the sequence of training sets in $\mathcal{S}$, one at a time. Starting with some initial parameters $\theta_0$. When a new training set $S_t$ is fed into $\ranker_{t-1} \coloneqq \ranker(\theta_{t-1})$, the parameter $\theta_{t-1}$ is updated to $\theta_t$ by learning the samples in $S_t$, yielding the new model $\ranker_t \coloneqq \ranker(\theta_t)$. 

\subsection{Performance evaluation} \label{section.evalutaion.a}

Let $\widetilde{\mathcal{S}} \coloneqq \{\widetilde{S}_1, \dotsc, \widetilde{S}_\nTasks\}$ be a (sequential) \textit{test} data set (for \gls{nir}). That is, for each $t \in \{1, \dotsc, T\}$,  $\widetilde{S}_t = \{(\tilde{x}_t^i, \tilde{y}_t^i)\}_{i=1}^{\tilde{n}_t}$, again contains a finite number of tuples of query--document pairs $\tilde{x}_t^i = (\tilde{q}_t^i, \tilde{d}_t^i) \in Q_t \times D_t$ and associated (ground-truth) relevance scores $\tilde{y}_t^i \in \mathbb{R}$.

The model performance may then be evaluated as follows. When training on the $t$th training set $S_t$ is completed, the updated model $\ranker_t$ is applied to the  query--document pairs $\tilde{x}_s^i = (\tilde{q}_s^i, \tilde{d}_s^i)$ in the $s$th test set, for all $s \in \{1, \dotsc, T\},$ to yield a predicted relevance score, $\ranker_t(\tilde{x}_s^i)$.

Finally, classical information retrieval metrics such as mean reciprocal recall or mean average precision then compute performance scores $\performance_{t,s}$ based on the true relevance scores $\tilde{y}_s^i$ and the predicted scores $\smash{\ranker_t(\tilde{x}_s^i)}$. The choice of evaluation metric is flexible and depends on the specific retrieval scenario. Some possible choices are discussed later.

\begin{figure}[ht]
\centerline{\includegraphics[scale=0.5]{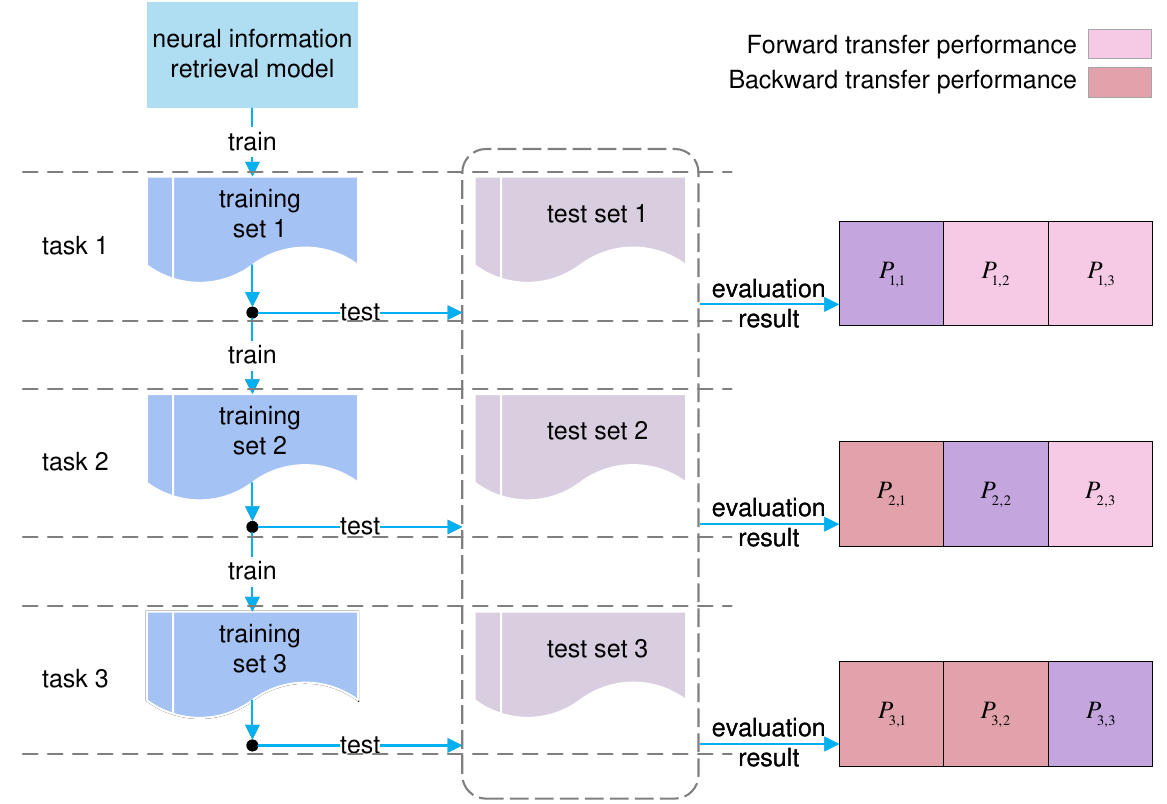}}

\caption{An example diagram of continual neural information retrieval with three tasks ($\nTasks = 3$). The neural information retrieval model is initially trained on the training set of task 1 and then tested on all three test sets to generate $\performance_{1,1}$, $\performance_{1,2}$, and $\performance_{1,3}$. After training on the set for task 2, $\performance_{2,1}$, $\performance_{2,2}$, and $\performance_{2,3}$ will be generated. Finally, upon completion of task 3, the model will produce $\performance_{3,1}$, $\performance_{3,2}$, and $\performance_{3,3}$.}
\label{Figure.0}
\end{figure}

Figure~\ref{Figure.0} provides an example of a continual neural information retrieval task. It illustrates how the model adapts to a continuously growing set of queries and documents and continually records its performance on multiple test sets over time. 

\subsection{Evaluation metrics}\label{section.evalutaion}

The average evaluation performance of $\ranker_\nTasks = \ranker(\theta_\nTasks)$ (the retrieval model after learning all tasks) is
 \begin{equation}
     \finalPerformance \coloneqq \frac{1}{\nTasks}\sum_{\taskIndex = 1}^{\nTasks} \performance_{\nTasks, \taskIndex}. \label{eq.fp}
 \end{equation}

To further explore the fine-grained continual learning performance, this study introduces \textit{\gls{bwt}} and \textit{\gls{fwt}} \citep{lopez2017gradient} into the information retrieval scenario. 

The backward transfer score, $\bwt$, measures the influence that learning a task has, on average, on the performance of previous tasks:
 \begin{equation}
     \bwt \coloneqq \frac{2}{\nTasks(\nTasks - 1)} \sum_{t = 2}^{\nTasks}\sum_{s = 1}^{t}(\performance_{t, s} - \performance_{s, s}). \label{eq.bwt}
 \end{equation}
A negative score, $\bwt < 0$, indicates the average performance reduction of the previous test sets on newly trained tasks. Conversely, $\bwt \geq 0$ indicates, on average, an absence of catastrophic forgetting.

The forward transfer score, $\fwt$, measures the influence that learning a task has on the performance of future tasks:
\begin{equation}
     \fwt \coloneqq \frac{2}{\nTasks (\nTasks - 1)} \sum_{t=1}^{\nTasks-1}\sum_{s=t+1}^{\nTasks} P_{t,s}.\label{eq.fwt}
\end{equation}
 A large value of $\fwt$ indicates that, on average, the model performs well on test sets of unlearned tasks; a small value of $\fwt$ indicates that the model is not able to effectively transfer knowledge from previous tasks to new ones.

\section{Proposed \gls{clnir} framework architecture} \label{section4}

\glsreset{clnir}

\begin{figure}[ht]
\centerline{\includegraphics[scale=0.6]{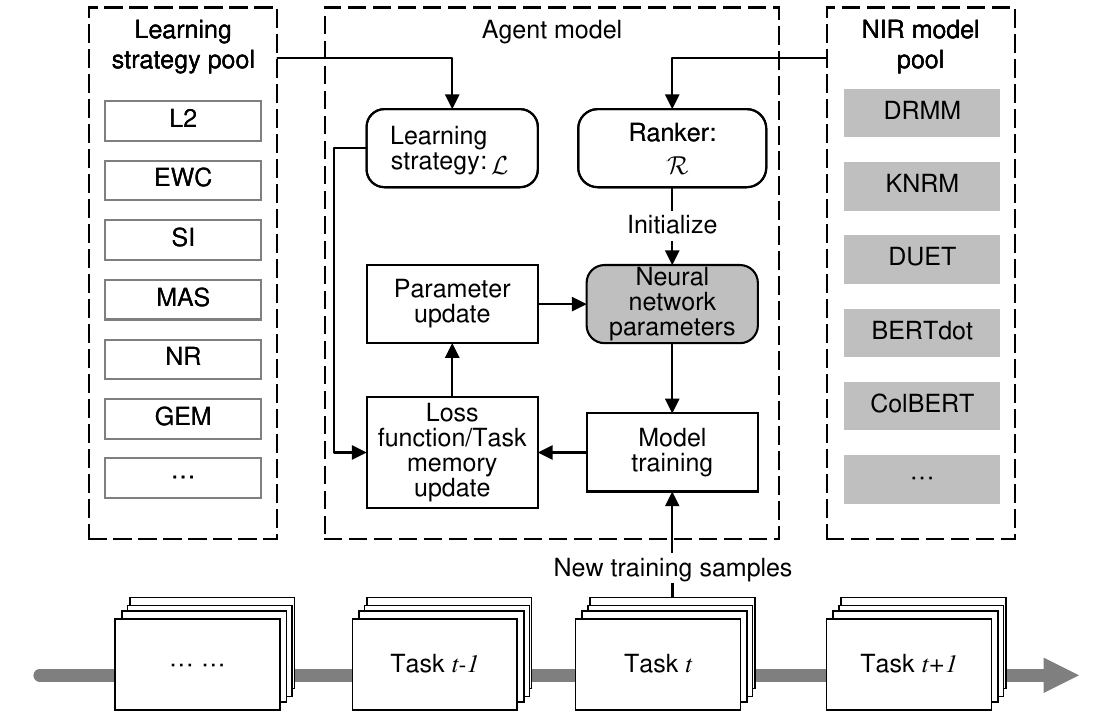}}
\caption{The \gls{clnir} framework architecture.The \gls{nir} models and learning strategies mentioned will be explained in Sections~\ref{section.nir_models} and \ref{section.learning_strategies} respectively.}
\label{Figure.4}
\end{figure}

This section introduces the proposed \textit{\gls{clnir}}. Inspired by a prior continual learning framework for classification tasks \citep{hsu2018re}, the \gls{clnir} framework allows for flexibility by decoupling the selection and implementation of the \gls{nir} model from the choice of learning strategy. The architecture of the proposed \gls{clnir} framework is depicted in Figure~\ref{Figure.4}. It consists of a learning strategy pool, a neural ranking model pool, and an agent model that performs \gls{nir} tasks.

\subsection{Agent model}

Algorithm~\ref{algo_clnir} shows the workflow of the \gls{clnir} framework. Line~1 initialises the agent model of the \gls{clnir} framework with a designated \gls{nir} model (ranker  $\ranker$), learning strategy ($\learningStrategy$) and a loss function ($\lossFunction$). The neural parameters within $\ranker$ are the optimization target of $\lossFunction$. 
When new training samples are fed into \gls{clnir}, $\ranker$ first learns the new data, similar to non-incremental learning tasks (Line~5). The \gls{clnir} framework allows for two types of learning strategies:
\begin{itemize}
    \item For \textit{regularization}-based strategies ($\learningStrategy \in \regularizationStrategySet$), Line~7 lets the agent model generate a new regularization item, $\regularizationFunction_t$, which measures the importance of the different components of $\theta_t$. Line~8 then updates the loss function to account for the modified regularization term. A number of suitable regularization-based strategies will be described in Section~\ref{strategies.regularization}. 
 
   \item For \textit{replay}-based strategies ($\learningStrategy \in \replayStrategySet$), Line~10 first initializes an empty memory set, $\memorySet$, when processing the first task. Line~11 then updates the memory set from the current training set using certain replay rules. For certain replay-based strategies, the loss function or the training set for the next task then need to be updated based on $\memorySet$ (Lines~12--13). A number of suitable replay-based strategies will be described in Section~\ref{strategies.replay}.

\end{itemize}

After the training process, Line~14 updates the agent model; Lines~15--18 then send all test sets into the updated agent model to generate performance scores. 
 
\IncMargin{1em}
\begin{algorithm} [htbp]
\SetKwFunction{Train}{Train}\SetKwFunction{Test}{Test}
\SetKwFunction{CalculateParameterImportance}{CalculateParameterImportance}
\SetKwFunction{UpdateTaskMemory}{UpdateTaskMemory}
\SetKwFunction{UpdateLossFunction}{UpdateLossFunction}
\SetKwFunction{UpdateNextTrainingData}{UpdateNextTrainingData}
\SetKwInOut{Input}{input}\SetKwInOut{Output}{output}
	\Input{continual \gls{nir} agent model $\agentModel$;\\ neural ranker $\ranker$ with initial parameter $\theta_0$;\\ learning strategy $\learningStrategy$;\\ loss function $\lossFunction$;\\ sequential \gls{nir} (training and test) datasets $\mathcal{S}$ and $\widetilde{\mathcal{S}}$ with $\nTasks$ tasks; }
	
     \Output{$\bwt$; $\fwt$; $\finalPerformance$;}
    \BlankLine 
	 initialize agent model: $\agentModel_{0}\leftarrow \agentModel(\ranker(\theta_0), \learningStrategy)$\; 
	 initialize $U$ as an empty list\; 
	 initialize $V$ as an empty list\; 
	 \For{$t \leftarrow 1$ \KwTo $\nTasks$}{
	 	$\theta_t \leftarrow$ \Train{$\agentModel_{t-1}$, $S_t$}\;
	 	\If()
	 	    {$\learningStrategy \in \regularizationStrategySet$}{

                 $\regularizationFunction_t \leftarrow $ \CalculateParameterImportance{$S_t, \theta_t,\regularizationFunction_{t-1}, \learningStrategy$}\;
	 	     $\lossFunction \leftarrow$ \UpdateLossFunction{$\lossFunction$, $\regularizationFunction_t$}\;
	 	    }
	 	\If()
	 	    {$\learningStrategy \in \replayStrategySet$}{\lIf{$t=0$}{initialize memory set $\memorySet$ an empty list} 
              $\memorySet\leftarrow$ \UpdateTaskMemory{$S_t$, $\memorySet$}\;
	 	 $\lossFunction \leftarrow$ \UpdateLossFunction{$\lossFunction$, $\memorySet$}\;
              $S_{t+1}\leftarrow$ \UpdateNextTrainingData{$S_t$, $\memorySet$}\;
	 	    }
    update agent model: $\agentModel_t \leftarrow \agentModel(\ranker(\theta_t), \learningStrategy)$\;
	 	\For{$s \leftarrow 1$ \KwTo $\nTasks$}{
	 	    $\performance_{t,s} \leftarrow $ \Test{$\agentModel_t$, $\widetilde{S}_s$}\;
	 	    \lIf{$s \leq t$}{append $\performance_{t,s}$ to $U$}
	 			 \lElse{append $\performance_{t, s}$ to $V$} 
	 	}
 	 }
 	 	\caption{Continual neural information retrieval}
 	 	\label{algo_clnir}
 	 \end{algorithm}
 \DecMargin{1em} 
 
\subsection{\gls{nir} model pool} \label{section.nir_models}

The proposed \gls{clnir} framework consists of three state-of-the-art \textit{embedding}-based methods and two \textit{pretraining}-based methods. However, researchers who adopt the proposed framework can include other models.

\subsubsection{Embedding based \gls{nir} models}

\begin{figure}[h]
		\begin{subfigure}{.3\textwidth}
			\centering
			\includegraphics[angle=90, scale=0.3]{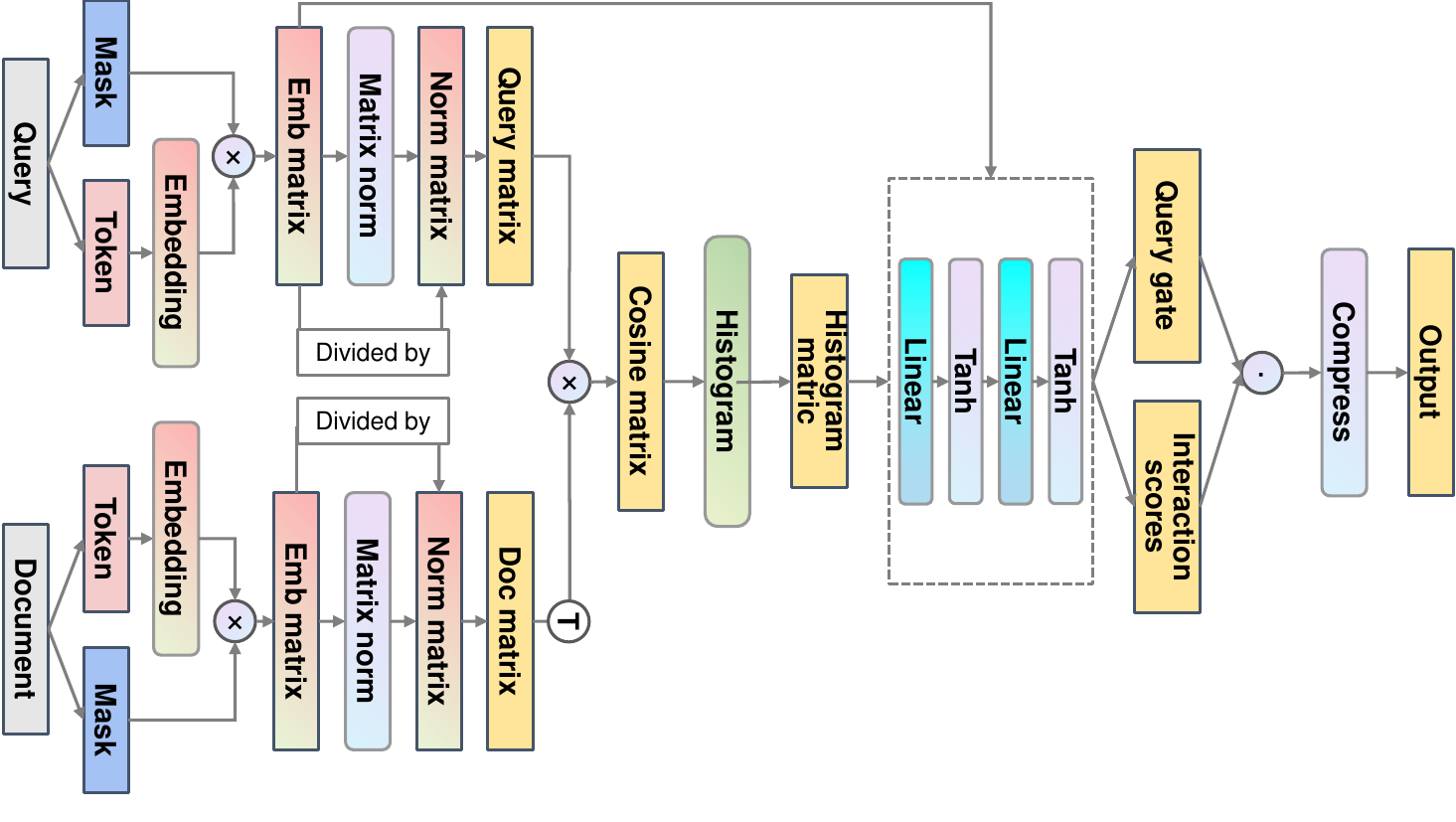} 
		\caption{\gls{drmm}}
     	\label{figure.5(a)}
		\end{subfigure}
        \hfil
		\begin{subfigure}{.3\textwidth}
			\centering
			\includegraphics[angle=90, scale=0.3]{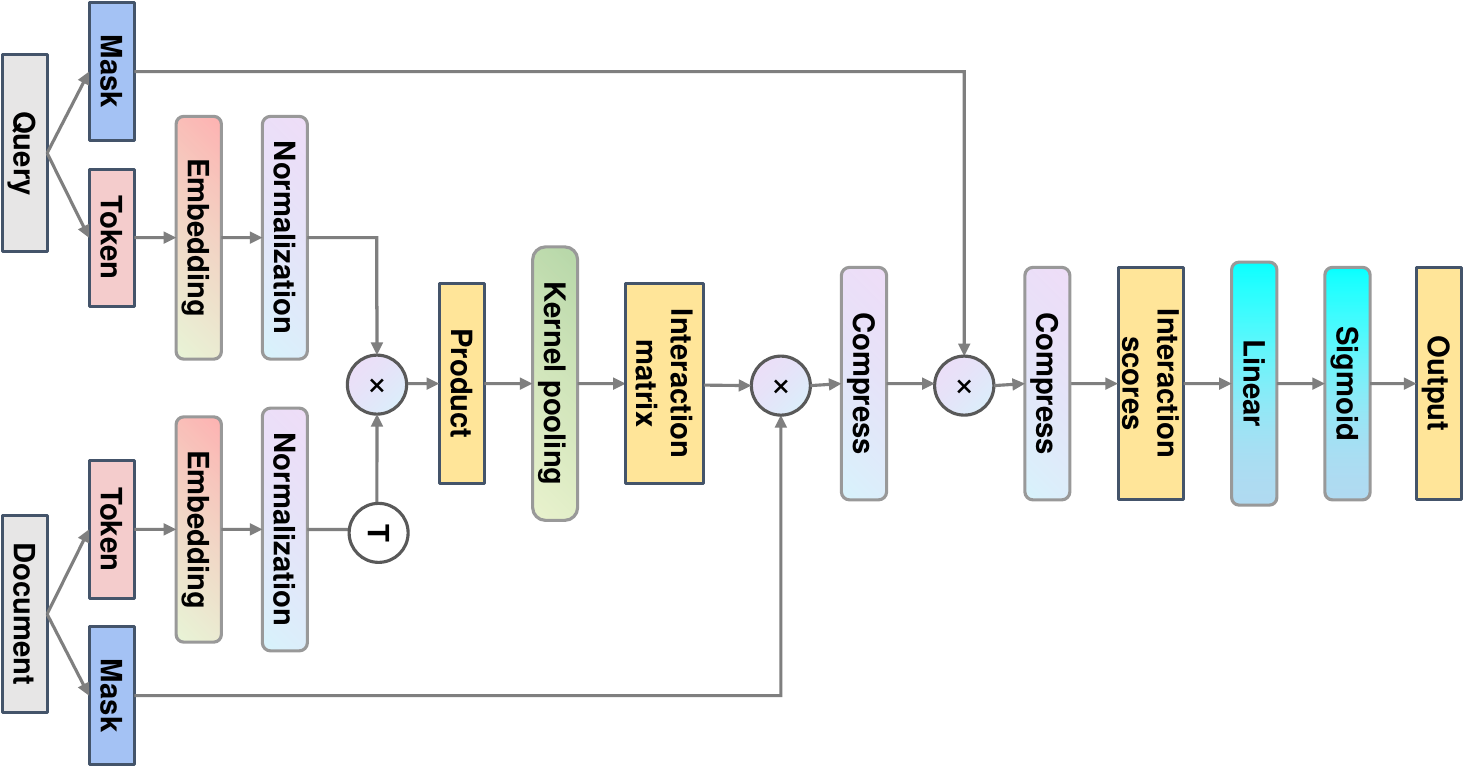}
			\caption{\gls{knrm}} 
            \label{figure.5(b)}
		\end{subfigure}
        \hfil
  	\begin{subfigure}{.3\textwidth}
			\centering
			\includegraphics[angle=90, scale=0.3]{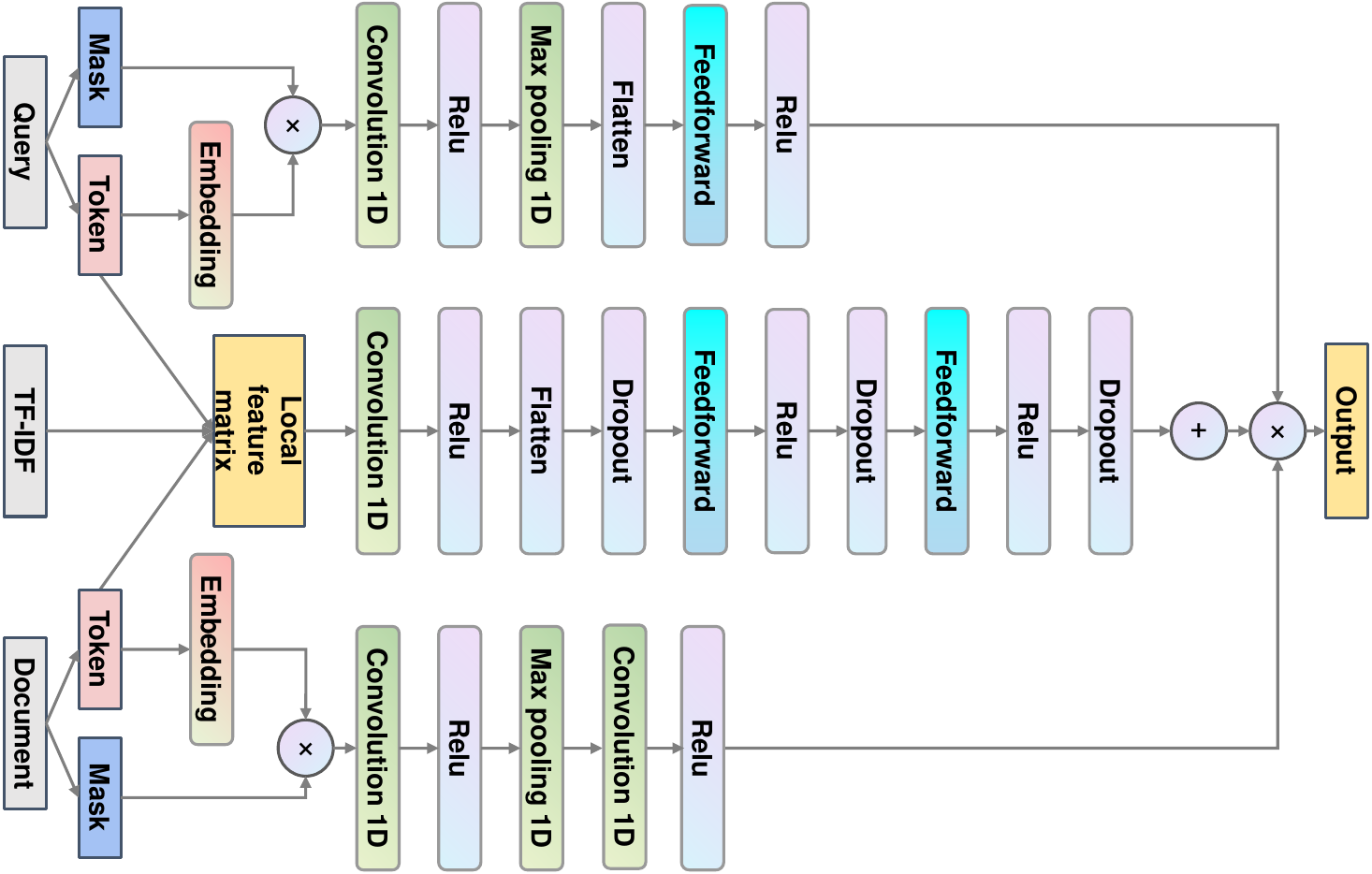}
			\caption{\gls{duet}} 
            \label{figure.5(c)}
		\end{subfigure}
  		\caption{Architectures of word embedding based \gls{nir} models.} 
	\end{figure}

 \textbf{DRMM}. \textit{\glspl{drmm}} \citep{guo2016deep} are interaction-based \gls{nir} models. Their architecture is depicted in Figure~\ref{figure.5(a)}. Given a query text $\singleQuery$ and a candidate document $\singleDocument$, the \gls{drmm} model tokenizes the texts into fixed-length tokens $\singleQueryTokens$ and $\singleDocumentTokens$, and generates corresponding masks $\singleQueryMask$ and $\singleDocumentMask$. The tokens are then input into an embedding layer, and the resulting embedding matrices are masked by $\singleQueryMask$ and $\singleDocumentMask$. The produced matrices for $\singleQuery$ and $\singleDocument$ can be represented as $\singleQueryEmbeddings \in \reals^{\queryLength \times \embeddingDimension}$ and $\singleDocumentEmbeddings \in \reals^{\documentLength \times n}$, respectively, where $\queryLength$ and $\documentLength$ represent the lengths of the query and document, respectively, and $n$ is the embedding dimension. The query representation is obtained by dividing $\singleQueryEmbeddings$ by its matrix norm, while the document representation is obtained from $\singleDocumentEmbeddings$ in the same manner. Both representation matrices have the same dimensions as their corresponding embedding matrices.

The relevance between query representation and document representation is determined based on the cosine matrix similarity and tensor histogram. The cosine matrix $M_{\mathrm{cos}} \in \reals^{\queryLength \times \documentLength}$ is obtained by taking the product of query representation and the transpose of document representation. Then, each row in $M_{\mathrm{cos}}$ is sorted into equal width ($b$) by a histogram function, resulting in a transformed matrix $M_{\mathrm{his}} \in \reals^{\queryLength \times b}$, where $b$ is the number of bins. Next, a series of linear and hyperbolic tangent (Tanh) layers are used to compress $M_{\mathrm{his}}$ and $\singleQueryEmbeddings$ into a relevance score vector $M_{\mathrm{rel}} \in \reals^{\queryLength}$ and a query gate vector $M_{\mathrm{gate}} \in \reals^{\queryLength}$, respectively. Finally, the relevance score is obtained by summing the element-wise product of $M_{\mathrm{rel}}$ and $M_{\mathrm{gate}}$.

\textbf{KNRM}. The architecture of \textit{\glspl{knrm}} \citep{xiong2017end} 
is shown in Figure~\ref{figure.5(b)}. Similar to \gls{drmm}, their input includes tokens $\singleQueryTokens$ and $\singleDocumentTokens$, and masks $\singleQueryMask$ and $\singleDocumentMask$. Then tokens $\singleQueryTokens$ and $\singleDocumentTokens$ are sent into an embedding layer to acquire query and document embeddings $\singleQueryEmbeddings$ and  $\singleDocumentEmbeddings$, which have the same shape as embedding matrices of \gls{drmm}. After normalization, $\singleQueryEmbeddings$ and  $\singleDocumentEmbeddings\tran \in \reals^{n \times \documentLength}$ are multiplied to an interaction matrix $M_{\mathrm{inter}} \in \reals^{\queryLength \times \documentLength}$.

The $k$-kernel pooling function is then applied to $M_{\mathrm{inter}}$ to create a kernel pooling tensor $M_{\mathrm{pool}} \in \reals^{\queryLength \times \documentLength \times k}$, where $k$ is the kernel size. The  $M_{\mathrm{pool}}$ is multiplied with $\singleDocumentMask$ to block unnecessary information, and then reduced to two dimensional $M'_{\mathrm{pool}} \in \reals^{\queryLength \times k}$ by summing the elements in the second dimension. The tensor $M'_{\mathrm{pool}}$ is then multiplied with $\singleQueryMask$ to block further unnecessary information and reduced to one dimensional $M''_{\mathrm{pool}} \in \reals^{k}$ by summing the elements in the first dimension. Finally, $M''_{\mathrm{pool}}$ is processed by a feed-forward and sigmoid layer to produce a final relevance score.

\textbf{DUET.} The \textit{Duet} \citep{mitra2017learning}  architecture consists of both a local model and a distributed model. The local model uses a local representation to match the query and document, while the distributed model uses learned distributed representations for the same purpose. The architecture of Duet is shown in Figure~\ref{figure.5(c)}. In contrast to \gls{drmm} and \gls{knrm}, Duet includes an additional term frequency-inverse document frequency (TF-IDF) module to initialize a local distribution model $M_{\mathrm{local}} \in \reals^{\queryLength \times \documentLength}$. Each element $w_{i}^{j} \in M_{\mathrm{local}}$ represents the TF-IDF score for the $i$th term in $\singleQuery$ and the $j$th word in $\singleDocument$ if the two terms are the same. The TF-IDF matrix is then flattened to one dimension, a vector $M'_{\mathrm{local}} \in \reals^{n}$ for the local representation.

In parallel, the embeddings of the query $\singleQueryEmbeddings \in \reals^{\queryLength \times n}$ and document $\singleDocumentEmbeddings \in \reals^{\documentLength\times n}$ undergo one-dimensional convolution, max-pooling, and other tensor operations to derive the query representation $M_{q} \in \reals^{n}$ and document representation $M_{d} \in \reals^{n \times c}$, respectively, where $c$ represents the output channel dimension. The product of $M_{q}$ and $M_{d}$ is then added to $M'_{\mathrm{local}}$ to generate a relevance score.

\subsubsection{Pretraining based \gls{nir} models}

\begin{figure}[h]
\begin{subfigure}{.45\textwidth}
    \centering
    \includegraphics[angle=90,scale=0.3]{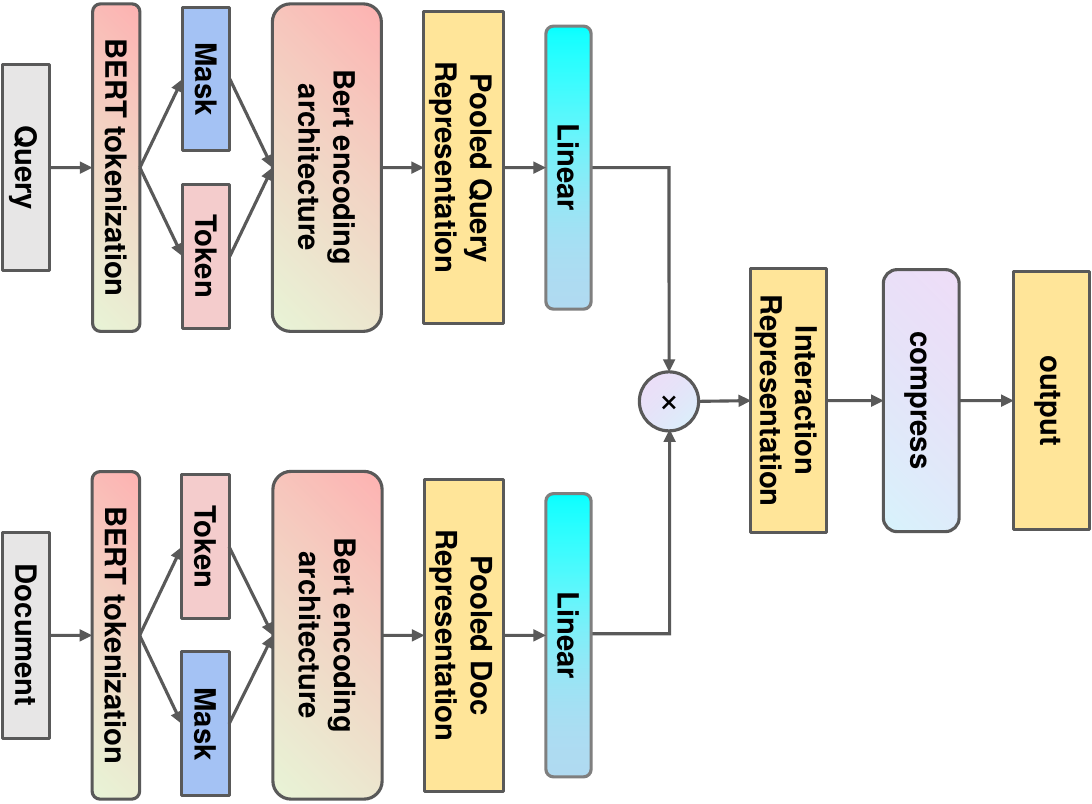}
    \caption{\gls{bertdot}}
    \label{figure.6(a)}
    \end{subfigure}
    \hfill 
\begin{subfigure}{.45\textwidth}
    \centering
    \includegraphics[angle=90, scale=0.3]{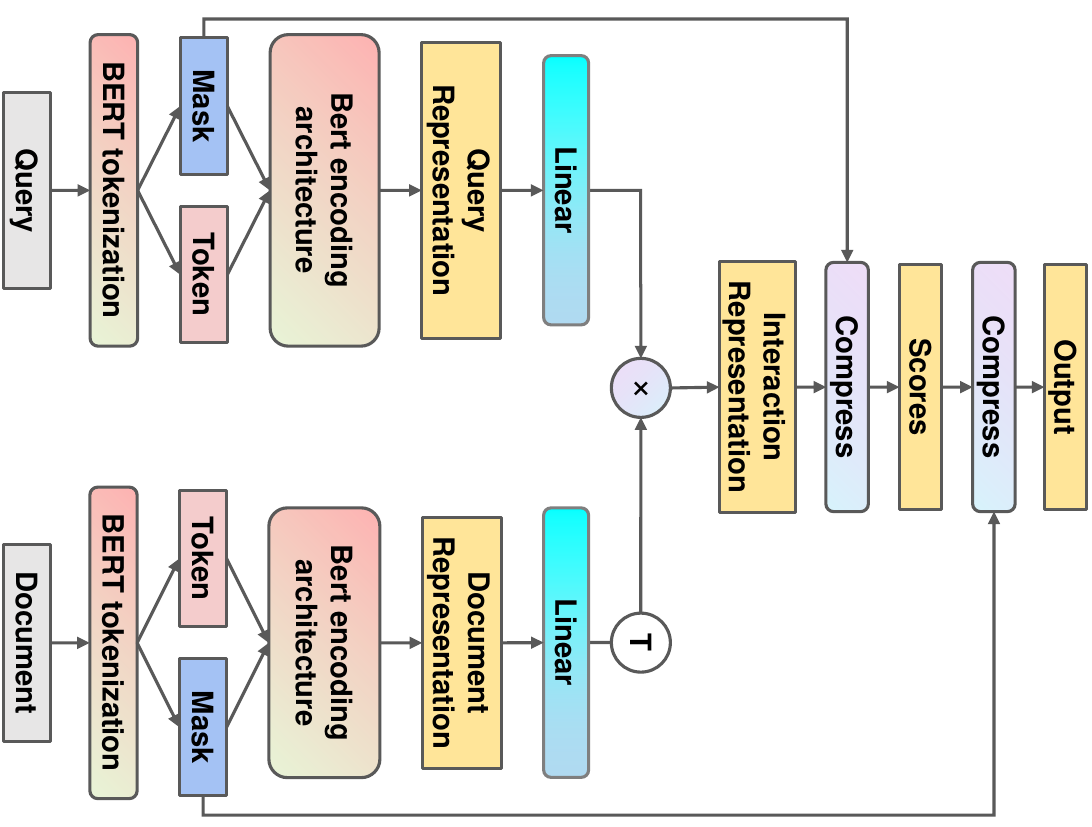}
    \caption{\gls{colbert}} 
    \label{figure.6(b)}
    \end{subfigure}
  \caption{Architectures of pre-trained \gls{nir} models.}
\end{figure}

\textbf{BERTdot.} \textit{\gls{bertdot}} \citep{hofstatter2021efficiently} is a direct application of the BERT model \citep{DBLP:conf/naacl/DevlinCLT19}. Its architecture is shown in Figure~\ref{figure.6(a)}. A BERT tokenizer processes the query text $\singleQuery$ and document text $\singleDocument$ to produce tokens and masks, which are then input into the BERT encoding layer to generate text representation vectors. \gls{bertdot} uses the pooled vectors for query and document representation, denoted as $\singleQueryBertEmbeddings \in \reals^{n}$ and  $\singleDocumentBertEmbeddings \in \reals^{n}$, where $n$ is typically set to 768, the default hidden size of BERT. To calculate the dot product between these representations, they are expanded to $\singleQueryBertEmbeddings \in \reals^{1 \times n}$ and $\singleDocumentBertEmbeddings \in \reals^{n \times 1}$. The resulting output vector is then condensed to a scalar to represent the relevance score.

\textbf{ColBERT.} The \textit{\gls{colbert}} model \cite{khattab2020colbert} encodes queries and documents separately and calculates the query-document similarity using a late interaction architecture, as shown in Figure~\ref{figure.6(b)}. In contrast to \gls{bertdot}, the query and document representations, $\singleQueryBertEmbeddings \in \reals^{\queryLength \times n}$ and $\singleDocumentBertEmbeddings \in \reals^{\documentLength \times n}$, respectively, are two-dimensional tensors encoded from the BERT architecture. The query representation and the transposed document representation are multiplied to obtain an interaction matrix $M_{\mathrm{inter}} \in \reals^{\queryLength \times \documentLength}$. The maximum value in the last dimension of $M_{\mathrm{inter}}$ is then taken to represent the query term similarity score $M_{q} \in \reals^{\queryLength}$. Finally, all the similarity scores are summed to generate the final relevance score.

\subsection{Learning strategies pool} \label{section.learning_strategies}
In a real-world retrieval scenario, the amount of emerging data is vast (potentially infinite), making it unrealistic to employ parameter isolation strategies that demand a separate parameter space for every new dataset. While replay-based methods also face this challenge, it is possible to establish a fixed memory space of adequate size. Regularization-based strategies, on the other hand, are generally not influenced by the number of tasks, therefore, the learning strategy pool considered in this study encompasses both the widely used regularization-based and replay-based methods.

\subsubsection{Regularization strategies for \gls{nir}} \label{strategies.regularization}


Let $\mathbf{x} =\pairwiseRankingSample$ be a batch of pairwise ranking samples for the current model input, where $\queryBatch$ represents a batch of queries, $\documentBatchPositive$ and $\documentBatchNegative$ represent either two document batches relevant and irrelevant to $\queryBatch$, respectively, or two documents where the former is more relevant to $\queryBatch$ than the latter. Regularization-based strategies aim to control parameter updates with a penalty item in the loss function. 
The resulting penalized loss function is: 
\begin{equation}
    \lossFunction(\mathbf{x};\theta)=\lossFunction'(\mathbf{x};\theta) + 
    \lambda\sum_{i}(\theta^i - \vartheta^{i})^{2}  \times \regularizationFunction^{i},
    \label{eq.cl_loss}
\end{equation}
where $\lambda > 0$ is the importance coefficient; $\vartheta = (\vartheta^i)$ is the current value of the parameter $\theta = (\theta^i)$; $\regularizationFunction = (\regularizationFunction^i)$ is a regularization item obtained from the previous task to measure the importance of the $i$th parameter in $\theta$. Then, to train $x$ pairwise, a margin ranking loss is used for loss calculation. That is, $\lossFunction'(\mathbf{x}; \theta)$ is the margin ranking loss function, defined as:
\begin{align}
  \lossFunction'(\mathbf{x};\theta) = \max\{0, -y (\ranker(\queryBatch, \documentBatchPositive; \theta) - \ranker(\queryBatch, \documentBatchNegative; \theta)) + \margin\}.
\label{eq.loss}
\end{align}
Here, $\ranker(\queryBatch, \documentBatchPositive; \theta)$ calculates the relevance scores between  $\queryBatch$ and $\documentBatchPositive$, while $\ranker(\queryBatch, \documentBatchNegative; \theta)$  calculates the scores between $\queryBatch$ and $\documentBatchNegative$; $\margin$ is a parameter to control the minimal difference between two ranking scores; and $y$ is a constant value. The difference between various regularization strategies lies in the addition of distinct regularization items in the loss function. The following are some methods used to implement these regularization items.

\textbf{L2.} The \gls{l2} strategy is a direct application of \gls{l2}-regularization. The parameter importance is set to a constant to keep every parameter equal in importance:
\begin{equation}
    \regularizationFunction^{i} = 1. \label{eq.l2}
\end{equation}

\glsreset{ewc}
\glsreset{ewcol}
\textbf{EWC and EWCol.} The \textit{\gls{ewc}} \cite{kirkpatrick2017overcoming} strategy calculates the parameter importance based on the parameter gradient with the Fisher information matrix. After all samples in a training set $S_t$ have been learned, a subset $S_t^* \subseteq S_t$ of $K$ training samples, called \textit{Fisher samples}, is chosen to build a so-called Fisher information matrix.
Then $S_t^*$ can be organized pairwise and the accumulated gradients generated from pairwise samples are used as parameter importance. The $k$th Fisher sample is denoted as $x_{k}^{*}=\langle q_{k}^{*},pos_{k}^{*},neg_{k}^{*} \rangle$. The average square of the gradients of each parameter is used as the regularization item for the next task. That is, if $\vartheta$ is the current parameter, set the regularisation item to
\begin{equation}
    \regularizationFunction^{i}=\frac{1}{K}\sum_{k=1}^{K}\mathbf{g}^i(x_k^*; \vartheta)^2,
    \label{eq.ewc}
\end{equation}
where
\begin{equation}
    \mathbf{g}^i(x; \vartheta) =\frac{\partial \lossFunction'(x;\vartheta)}{\partial\vartheta^i}.
\end{equation}
For \textit{\gls{ewcol}},  the previously accumulated importance factors will be added to the new ones: 
\begin{equation}
    \regularizationFunction^{i} \leftarrow \regularizationFunction^{i}+ \frac{1}{K}\sum_{k=1}^{K}
    \mathbf{g}^i(x_k; \vartheta)^2.
    \label{eq.ewc_online}
\end{equation}

\glsreset{si}
\textbf{SI}. The \textit{\gls{si}} \cite{zenke2017continual} strategy uses an importance measure to reflect the past credits for improvements of the regularized loss function for each parameter. Before training on a new batch $\mathbf{x}_k$, given the current parameter value $\vartheta_{k-1}$, with the convention that $\vartheta_0 = \vartheta$ is the parameter learned at the end of the previous task and $\vartheta_K = \theta$ is the new parameter at the end of the current task. Then \gls{si} updates the parameter from $\vartheta_{k-1}$ to $\vartheta_{k}$ by learning $\mathbf{x}_k$ with a regularized loss function. An importance measure is used to accumulate the importance of each parameter, which is calculated as the product of the gradient and the distance between the two parameters:
  \begin{equation}
    \omega^{i} = -\sum_{k} \mathbf{g}^i(x_k;\vartheta_{k-1}) \times (\vartheta_k^{i}-\vartheta_{k-1}^{i}).\label{eq.si.w}
\end{equation}
The regularization item is then updated as
\begin{equation}
\regularizationFunction^{i} \leftarrow \regularizationFunction^{i} + \frac{\omega^{i}}{(\theta^{i}-\vartheta^{i})^2+\xi }, \label{eq.si}
\end{equation}
where $\xi$ is a damping factor to avoid denominators becoming zero. It should be noted that $\vartheta = \vartheta_0$ and $\theta = \vartheta_K$ are the parameters at the end of the previous and current tasks, respectively.

\glsreset{mas}
\textbf{MAS}. The \textit{\gls{mas}} \cite{aljundi2018memory} computes the importance of each parameter in the network by measuring the sensitivity of the predicted output function to changes in that parameter. After completing the training process for the previous task, the resulting parameter $\theta$ is obtained. Before starting the next training process, the \gls{mas} method reuses all the pairwise training samples and accumulates the gradients of the square difference of relevance scores in each pair with respect to $\theta$. For the $k$th batch, $\mathbf{x}_k = \langle\queryBatch_k, \documentBatchPositive_k, \documentBatchNegative_k \rangle$ in the current task, the process can be denoted as follows:

\begin{equation}
    \regularizationFunction^{i} \leftarrow \regularizationFunction^{i} + \frac{1}{K}\sum_{k=1}^K  \biggl\lvert \frac{\partial (\ranker(\queryBatch_k, \documentBatchPositive_k; \theta)-\ranker(\queryBatch_k, \documentBatchNegative_k; \theta))^2}{\partial \theta^i} \biggr\rvert.
    \label{eq.max} 
\end{equation}

\subsubsection{Replay strategies for \gls{nir}} \label{strategies.replay}
\textbf{NR.} The \textit{\gls{nr}} is a direct replay method. Before training a new task, previously accumulated samples will be added to the new training set.  The new training set, $\mathbf{W}_{t}$ is a union of the current training set and the previous memory sets: 
\begin{align}
    \mathbf{W}_{t} \coloneqq S_t \cup \memorySet, \quad \text{with} \quad 
    \memorySet \coloneqq \bigcup_{s=1}^{t} \memorySet_{s}^{t-1}, \label{eq.replay.data} 
\end{align}
where $\memorySet_s^t$ represents a set of samples from the $s$th task that are used for training in the $t$th task.

To ensure that the total memory size, $\# \bigcup_{s=1}^t \memorySet_s^{t-1}$ remains consistent across different tasks, samples are regularly removed so that $\memorySet_s^1 \supseteq  \memorySet_s^2 \supseteq \dotsb \supseteq \memorySet_s^{t-1}$.

\glsreset{gem}
\textbf{GEM.} The \textit{\gls{gem}} \cite{lopez2017gradient} strategy still uses a memory set to alleviate catastrophic forgetting. Unlike \gls{nr}, the memory set in \gls{gem} is used for the update of the gradient for new task samples. Before training a new batch $\mathbf{x}_k$, \gls{gem} calculates the gradients of all memory samples with respect to the current parameter value $\vartheta_{k-1}$ in a set-wise manner.
 
 Let $\mathbf{G}_\mathbf{M}=(\mathbf{g}_\mathbf{M}^1,\mathbf{g}_\mathbf{M}^2,\dotsc,\mathbf{g}_\mathbf{M}^{t-1})\tran$, where $\mathbf{g}_\mathbf{M}^s$ is the gradient of the $s$th memory set. Then $\mathbf{x}_k$ will be learned and the corresponding parameter is updated to $\vartheta_k$, and $\mathbf{g}_{k}$ represents the new gradient.

The angle between $\mathbf{g}_{k}$ and all memory gradients will be used as a constraint to control parameter update, shown as:
\begin{equation*}
     \mathbf{g}_{k}\tran\mathbf{g}_\mathbf{M}^s  \geq 0, \text{ for all } s \textless t. \label{eq.gem.constraint} 
\end{equation*}
If all constraints are satisfied, the $\mathbf{g}_{k}$ will continue being used as the gradient for the next sample. However, if constraints are violated, a gradient projection will be used to alleviate forgetting. The new gradient $\widetilde{\mathbf{g}_{k}}$ replaces $\mathbf{g}_{k}$:
\begin{equation}
    \widetilde{\mathbf{g}_{k}}=\mathbf{g}_\mathbf{M}\tran \times v^* +\mathbf{g}_{k}, \label{eq.gem.project} 
\end{equation}
where $v^*$ is the vector of the result of the dual quadratic program:
\begin{equation*}
\begin{aligned}
& \underset{v}{\text{minimize}}
& &  \frac{1}{2}v\tran\mathbf{g}_\mathbf{M}\mathbf{g}_\mathbf{M}\tran v + \mathbf{g}_{k}\tran\mathbf{g}_\mathbf{M}\tran v \\
& \text{subject to}
& & v \geq 0.
\end{aligned}
\end{equation*}

\section{Experiments} \label{section5} 

In the absence of a benchmark dataset tailored for evaluating continual learning in neural information retrieval, this section is initiated with dataset preparation, followed by the description of the experiment setup. Subsequently, the results of the evaluation of continual learning performance are presented, alongside an examination of the impact of data volume and topic shift. Finally, the key findings are summarized.

\subsection{Topic-MSMARCO: a new dataset for continual \gls{nir}}

To assess the performance of methods for continual \gls{nir}, a \gls{nir} dataset with multiple subsets that simulates the emergence of real-world data is necessary. MSMARCO~\cite{nguyen2016ms}, one of the most important and largest benchmark datasets for \gls{nir}, was selected to derive several subsets based on the topic distribution. The `Passage Ranking' dataset of MSMARCO is an information retrieval dataset comprising real-world questions and answers from a search engine.

\begin{figure}[htbp]
\centering
\includegraphics[width=0.8\textwidth]{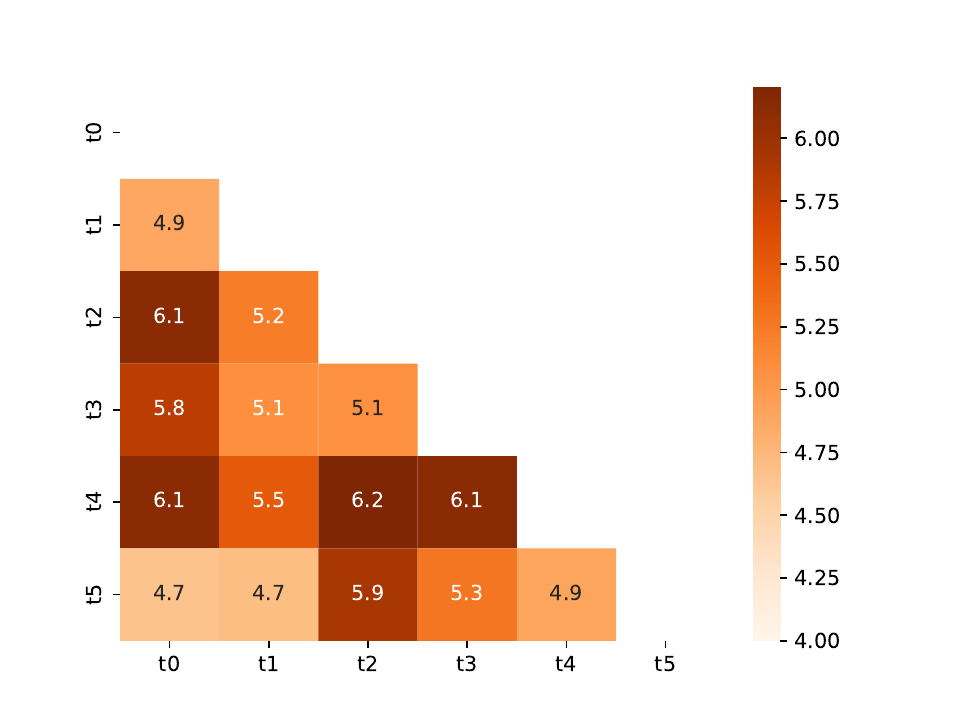}
\caption{Pairwise task distances in \textit{Topic-MSMARCO}. A larger value denotes a larger semantic distance.}
\label{figure.3(b)}
\end{figure}

A recent study~\cite{hong2021comparing} concluded that using a combination of word2vector and clustering models results in improved clustering outcomes. This methodology was applied in the study. In this study, only query texts were utilized to derive the topic distribution. The word2vector model was employed to train word embeddings on the cleaned query corpus. By incorporating the embedded query terms, a sequence of clustering experiments using the KMeans algorithm was executed, encompassing diverse cluster numbers. The optimal cluster configuration was then manually selected based on perplexity evaluations. Subsequently, any noisy clusters were expunged from consideration, and clusters exhibiting analogous thematic content were amalgamated to enhance coherence and interpretability. The created dataset is referred to as \textit{Topic-MSMARCO}. As shown in Table~\ref{table.1}, it encompasses six tasks (t0 to t5), each with a distinct topic. The pairwise topic distances (see Figure~\ref{figure.3(b)}) are defined as the squared distances between the respective cluster centres.

\begin{table}[htbp]
 \caption{Metadata and representative query terms of Topic-MSMARCO.}
\centering
\resizebox{\textwidth}{!}{\begin{tabular}{@{}llllll@{}}
\toprule
Topic index & Number of queries & Theme & Representative query terms &\\ \midrule

t0 & 64,719 & IT &
\begin{tabular}[c]{@{}l@{}}upload, emails, gmail, ios, icloud, cisco, photo,\\  powerpoint, audio, instagram\end{tabular} \\
 t1 & 40,501 & Furnishing &
 \begin{tabular}[c]{@{}l@{}}doors, fiberglass, cement, drywall, fireplace,\\  siding, stair, dryer, porch, ceramic\end{tabular} \\
 t2 & 51,002 & Food & \begin{tabular}[c]{@{}l@{}}broccoli, cabbage, canned, hamburger, fried,\\ carrots, greens, peas, dried, tuna,\end{tabular}\\
t3 & 47,646 & Health & \begin{tabular}[c]{@{}l@{}}painful, sleeping, vomiting, ache, weak, coughing,\\ canine, uterus, lip, dysfunction, thrush, chronic, \\ osteoarthritis, cell, mitochondria, subcutaneous,\\  narrow, forehead,\end{tabular} \\
t4 & 12,204 & Tourism & \begin{tabular}[c]{@{}l@{}}tampa, nashville, philadelphia, mall, downtown, rio,\\  denver, veniceniagara, cape, hills metro, zoo, \\ city, yellowstone\end{tabular}\\
t5 & 46,326 & Finance & \begin{tabular}[c]{@{}l@{}}employers, qualified,  orders, settlement, fraud,\\ resident, reimbursement, charges, pays, residency,\\ commission, officers\end{tabular}
\\ \bottomrule
\end{tabular}}
\label{table.1}
\end{table}

\subsection{Experimental setup} 

\textbf{Single \gls{nir} task evaluation.}
As described in the last paragraph in Section~\ref{section3.1}, different \gls{nir} tasks require different evaluation metrics. In this work, the performance score on a single test set is calculated by \textit{\gls{mrr}}, as it is an officially recommended metric from the MSMARCO team\footnote{https://microsoft.github.io/msmarco/}. If $Q$ is the set of queries (for which the test set includes relevance scores), the \gls{mrr} is defined as:
\begin{equation}
    \mathit{MRR}=\frac{1}{\#Q} \sum_{q \in Q} \frac{1}{rank_q}, \label{eq.mrr}
\end{equation}
where $\mathrm{rank}_q$ is the rank of the first relevant item for query $q$; $\#Q$ is the number of queries in $Q$. 

\textbf{Hardware and implementation.}
The hardware environment for the experiments was an Nvidia RTX 3080 GPU workstation with an Intel I7-10\,700K CPU and 64GB of RAM. All neural network models in \gls{clnir} were implemented using PyTorch\footnote{https://pytorch.org/}. The word embedding vectors used were GloVe \cite{pennington2014glove}, and 300-dimensional word vectors were used as the initial weight for the embedding layer of \gls{drmm}, \gls{knrm}, and \gls{duet}. The \gls{plm} used for \gls{colbert} and \gls{bertdot} is described in \citet{wolf2020transformers}.

\subsection{Evaluating \gls{clnir} using the topic-MSMARCO dataset}

The first experiment aims to compare the continual learning performance of various combinations of neural retrieval models and learning strategies within the  \gls{clnir} framework proposed in Section~\ref{section4}. 

To eliminate the potential impact of inconsistent sample sizes in each task, $10\,000$ queries were randomly selected as the training set and $2\,000$ queries were randomly selected as the test set for each task. The \gls{clnir} framework was used to pair neural ranking models with learning strategies, resulting in 35 continual retrieval methods. Five fine-tuning methods (without a learning strategy) were considered as baselines.

\subsubsection{Average final performance}

Table~\ref{table.2} shows the final performance score for the final average performance, $\finalPerformance$, for each possible combination of \gls{nir} model and learning strategy under the \gls{clnir} framework as well as for the baseline methods.

\begin{table}[htbp] \scriptsize
\caption{Final performance scores $\finalPerformance$ defined in Section~\ref{section.evalutaion} ($\pm$ standard error). Best performing strategies for each model are highlighted in bold.}
\centering
\begin{tabular}{@{}@{\hspace{5pt}}lrrrrr@{}}
\toprule
 \multicolumn{1}{c}{\multirow{2}{*}{{Strategy}}} & \multicolumn{5}{c}{Model}\\\cmidrule(l{1\cmidrulekern}){2-6}
 & {\gls{drmm}} & {\gls{knrm}} &  {\gls{duet}} & {\gls{bertdot}} &  {\gls{colbert}} \\ \midrule
Baseline & 0.074±0.002 & 0.173±0.003 & 0.129±0.002 & 0.237±0.003 & 0.322±0.003 \\
\gls{l2} & {\B 0.086±0.002} & {\B 0.185±0.003} & 0.147±0.002 & 0.240±0.003 & 0.331±0.003 \\
\gls{ewc} & 0.082±0.002 & {\B 0.185±0.003} & {\B 0.221±0.003} & 0.227±0.003 & 0.300±0.003 \\
\gls{ewcol} & {\B 0.085±0.002} & {\B 0.185±0.003} & {\B 0.222±0.003} & 0.210±0.003 & 0.295±0.003 \\
\gls{si} & 0.080±0.002 & {\B 0.185±0.003} & {\B 0.223±0.003} & 0.228±0.003 & 0.299±0.003 \\
\gls{mas} & 0.080±0.002 & {\B 0.183±0.003} & 0.181±0.003 & 0.244±0.003 & 0.325±0.003 \\
\gls{nr} & 0.081±0.002 & 0.174±0.003 & 0.186±0.003 & {\B 0.265±0.003} & {\B 0.346±0.003} \\
\gls{gem} & 0.061±0.002 & 0.134±0.002 & 0.144±0.002 & 0.241±0.003 & 0.312±0.003\\ \bottomrule
\end{tabular}%
\label{table.2}
\end{table}

\textbf{\gls{drmm} group.} Despite the lower \gls{mrr} scores compared with other \gls{nir} models, some continual learning strategies lead to performance improvement. Without employing any learning strategy, the final performance score is relatively low. However, with the \gls{l2} and \gls{ewcol}, it increases noticeably. The \gls{ewc}, \gls{mas}, \gls{nr}, and \gls{si} strategies also have a positive effect on the final performance. However, the \gls{gem} strategy does not offer any performance improvement in the \gls{drmm} group, leading to a lower final performance compared to the baseline.

\textbf{\gls{knrm} group.} The \gls{l2}, \gls{ewc}, \gls{ewcol}, \gls{si}, and \gls{mas} strategies yield higher scores, while the \gls{nr} strategy performs similarly to the baseline. The sole strategy that adversely affects the average final performance is \gls{gem}, which is even lower than the baseline.

\textbf{\gls{duet} group.} Continual learning strategies exhibit the most significant improvement in the \gls{duet} group. Notably, this group is the only one where all pair combinations of \gls{nir} models with various learning strategies outperform the baseline method. Particularly, the average final performance of \gls{ewc}, \gls{ewcol}, and \gls{si} represents the largest improvement among all \gls{nir} groups.

\textbf{\gls{bertdot} group.} Both replay strategies (\gls{nr}, \gls{gem}) outperform the baseline method. The highest score in this group is achieved by \gls{nr}, which is 0.028 higher than the baseline. It is worth noting that among all the regularization-based strategies, only \gls{mas} and \gls{l2} show a slight improvement over the baseline score, while the two \gls{ewc} strategies and \gls{si} result in lower scores compared to the baseline.

\textbf{\gls{colbert} group.} In this experiment, \gls{colbert} outperforms the other models. All strategies except for \gls{ewcol} yield an average final performance score higher than 0.3. Similar to the \gls{bertdot} group, \gls{nr} achieves the highest score. Both \gls{ewc} strategies and \gls{gem} demonstrate inferior performance compared to the baseline, while \gls{l2} slightly improves upon the baseline.

\subsubsection{Backward-transfer performance}

\textbf{\gls{bwt} performance of baseline methods.} As demonstrated in Table~\ref{table.3}, all the baseline methods experienced catastrophic forgetting as all their \gls{bwt} scores are negative. The \gls{bwt} scores of the \gls{duet} and \gls{bertdot} models are a few hundredths below zero, indicating that they are more susceptible to forgetting without learning strategies, compared to the remaining \gls{nir} models whose \gls{bwt} scores are a few thousandths under zero.

\textbf{Impact of learning strategies on \gls{bwt}.} As shown in Table~\ref{table.3}, when combined with learning strategies, all \gls{nir} models achieved nonnegative \gls{bwt} scores, indicating that \gls{nir} models can perform better on \gls{bwt} with learning strategies. The \gls{duet} model achieved the highest \gls{bwt} scores with the \gls{ewc}, \gls{ewcol} and \gls{si} strategies, at around 0.02. \gls{colbert} showed a higher degree of adaptability to diverse learning strategies, with only one of its \gls{bwt} scores less than zero when combined with learning strategies.

\textbf{Learning strategies comparison on \gls{bwt}.} 
With regards to the learning strategies, according to Table~\ref{table.3}, the \gls{si} strategy is observed to achieve the most positive \gls{bwt} scores, outperforming the other strategies. Both the \gls{mas} and \gls{nr} strategies have two positive \gls{bwt} scores on two pretraining-based models. In contrast, the \gls{l2} and two \gls{ewc} strategies yield only one positive score each. Notably, the \gls{gem} strategy stands out as the sole strategy unable to overcome catastrophic forgetting.

\begin{table}[htbp]\scriptsize
\caption{\gls{bwt} scores $\bwt$ defined in Section~\ref{section.evalutaion} ($\pm$ standard error). Best performing strategies for each model are highlighted in bold.}
\centering
\begin{tabular}{@{}@{\hspace{5pt}}lrrrrr@{}}
\toprule
 \multicolumn{1}{c}{\multirow{2}{*}{{Strategy}}} & \multicolumn{5}{c}{Model}\\\cmidrule(l{1\cmidrulekern}){2-6}
 & {\gls{drmm}} & {\gls{knrm}} &  {\gls{duet}} & {\gls{bertdot}} &  {\gls{colbert}} \\ \midrule
Baseline & -0.009±0.001 & -0.007±0.001 & -0.037±0.002 & -0.012±0.002 & -0.002±0.002 \\
\gls{l2} & -0.003±0.001 & {\B 0.000±0.000} & -0.020±0.002 & -0.010±0.002 & 0.004±0.002\\
\gls{ewc} & -0.001±0.001 & {\B 0.000±0.000} & {\B 0.018±0.001} & 0.000±0.000 & 0.000±0.000  \\
\gls{ewcol} & {\B 0.000±0.000} & {\B 0.000±0.000} & {\B 0.019±0.001} & -0.013±0.001 & 0.000±0.000 \\
\gls{si} & {\B 0.000±0.000} & {\B 0.000±0.000} & {\B 0.017±0.001} & 0.001±0.000 & 0.001±0.000\\
\gls{mas} & -0.007±0.001 & -0.002±0.001 & -0.006±0.002 & 0.001±0.000 & 0.003±0.002 \\
\gls{nr} &-0.004±0.001 & -0.007±0.001 & -0.004±0.002 & {\B 0.008±0.002} & {\B 0.008±0.002} \\
\gls{gem} & -0.013±0.001 & -0.028±0.001 & -0.032±0.002 & -0.010±0.002 & -0.006±0.002\\ \bottomrule
\end{tabular}%
\label{table.3}
\end{table}
\subsubsection{Forward-transfer performance}

\textbf{Impact of learning strategies on \gls{nir} models.}
Table~\ref{table.4} indicates that the \gls{fwt} scores of the \gls{nir} groups that pretraining-based \gls{nir} models outperform the embedding-based ones, consistent with their average final performance. \gls{colbert} paired with the \gls{nr} strategy achieves the highest \gls{fwt} score. Similarly, the combination of \gls{bertdot} and \gls{nr} also demonstrates notable performance within the \gls{bertdot} group. In the \gls{duet} group, the \gls{ewcol} strategy exhibits the most effective \gls{fwt} performance, followed closely by \gls{ewc} and \gls{si}. Within the \gls{knrm} and \gls{drmm} groups, all strategies except for \gls{gem} either maintain parity with or slightly enhance the \gls{fwt} performance compared to the baseline strategy.

\textbf{Comparison of \gls{fwt} between baselines and learning strategies.}
According to Table~\ref{table.4}, among all learning strategies, \gls{gem} stands out as the sole strategy that consistently reduces baseline \gls{fwt} scores to varying degrees in both the \gls{drmm} and \gls{knrm} groups. Notably, within the groups featuring pretraining-based \gls{nir} models, several learning strategies exhibit a negative impact on \gls{fwt} performance. Specifically, in the \gls{bertdot} group, two strategies (\gls{ewc} and \gls{si}) yield lower \gls{fwt} scores compared to the baseline, while in the \gls{colbert} group, the \gls{nr} strategy stands out as the only one surpassing the baseline method.

\begin{table}[htbp] \scriptsize
\caption{\gls{fwt} scores $\fwt$ defined in Section~\ref{section.evalutaion} ($\pm$ standard error). Best performing strategies for each model are highlighted in bold.}
\centering
\begin{tabular}{@{}@{\hspace{5pt}}lrrrrr@{}}
\toprule
 \multicolumn{1}{c}{\multirow{2}{*}{{Strategy}}} & \multicolumn{5}{c}{Model}\\\cmidrule(l{1\cmidrulekern}){2-6}
 & {\gls{drmm}} & {\gls{knrm}} &  {\gls{duet}} & {\gls{bertdot}} &  {\gls{colbert}} \\ \midrule
Baseline & 0.070±0.001 & 0.160±0.002 & 0.142±0.002 & 0.217±0.002 & 0.292±0.002 \\
\gls{l2} & {\B 0.074±0.001} & {\B 0.164±0.002} & 0.147±0.002 & 0.216±0.002 & 0.289±0.002  \\
\gls{ewc} & {\B 0.073±0.001} & {\B 0.164±0.002} & {\B 0.165±0.002} & 0.206±0.002 & 0.275±0.002 \\
\gls{ewcol} & {\B 0.074±0.001} & {\B 0.164±0.002} & {\B 0.167±0.002} & 0.222±0.002 & 0.271±0.002 \\
\gls{si} & 0.070±0.001 & {\B 0.164±0.002} & {\B 0.165±0.002} & 0.206±0.002 & 0.274±0.002\\
\gls{mas} & 0.072±0.001 & {\B 0.162±0.002} & 0.154±0.002 & 0.223±0.002 & 0.289±0.002 \\
\gls{nr} & {\B 0.074±0.001} & 0.160±0.002 & 0.153±0.002 & {\B 0.226±0.002} & {\B 0.300±0.002}\\
\gls{gem} & 0.063±0.001 & 0.156±0.002 & 0.151±0.002 & 0.218±0.002 & 0.287±0.002\\ \bottomrule
\end{tabular}%
\label{table.4}
\end{table}

\subsubsection{Overall comparison}

\textbf{Most effective learning strategies for evaluating Topic-MSMARCO.} From the previous comparisons of the final average performance, \gls{bwt} and \gls{fwt}, it can be observed that all five \gls{nir} models benefit from learning strategies. \gls{nir} models combined with the appropriate learning strategies always outperform the baseline methods in any evaluation metric. Notably, the two pretraining-based \gls{nir} models exhibit significant performance enhancement when paired with the \gls{nr} strategy, showcasing the most effective performance across all three metrics within their respective groups. However, the optimal learning strategies for different embedding-based \gls{nir} models vary. Detailed insights into the most effective learning strategies for different \gls{nir} models and evaluation metrics are provided in Table~\ref{table.5}.

\begin{table}[htbp]
\scriptsize
\caption{Most effective learning strategies for different \gls{nir} models on different evaluation metrics.}
\resizebox{\textwidth}{!}{
\centering
\begin{tabular}{@{}lccccc@{}}
\toprule
 \multicolumn{1}{c}{\multirow{2}{*}{{Metric}}} & \multicolumn{5}{c}{Model}\\\cmidrule(l{1\cmidrulekern}){2-6}
 & {\gls{drmm}} & {\gls{knrm}} &  {\gls{duet}} & {\gls{bertdot}} &  {\gls{colbert}} \\ \midrule
$\finalPerformance$ & \gls{l2}/\gls{ewcol} & \gls{l2}/\gls{ewc}/\gls{ewcol}/\gls{si}/\gls{mas} & \gls{ewc}/\gls{ewcol}/\gls{si} & \gls{nr}  & \gls{nr}\\
$\bwt$ & \gls{ewcol}/\gls{si} & \gls{l2}/\gls{ewc}/\gls{ewcol}/\gls{si} & \gls{ewc}/\gls{ewcol}/\gls{si} & \gls{nr}  & \gls{nr}\\
$\fwt$ & \gls{l2}/\gls{ewc}/\gls{ewcol}/\gls{nr} & \gls{l2}/\gls{ewc}/\gls{ewcol}/\gls{si}/\gls{mas} & \gls{ewc}/\gls{ewcol}/\gls{si} & \gls{nr}  & \gls{nr}\\
\bottomrule
\end{tabular}%
}
\label{table.5}
\end{table}

\begin{figure}[htbp]
	\begin{subfigure}{0.32\textwidth}
		\centering
		\includegraphics[width=\textwidth]{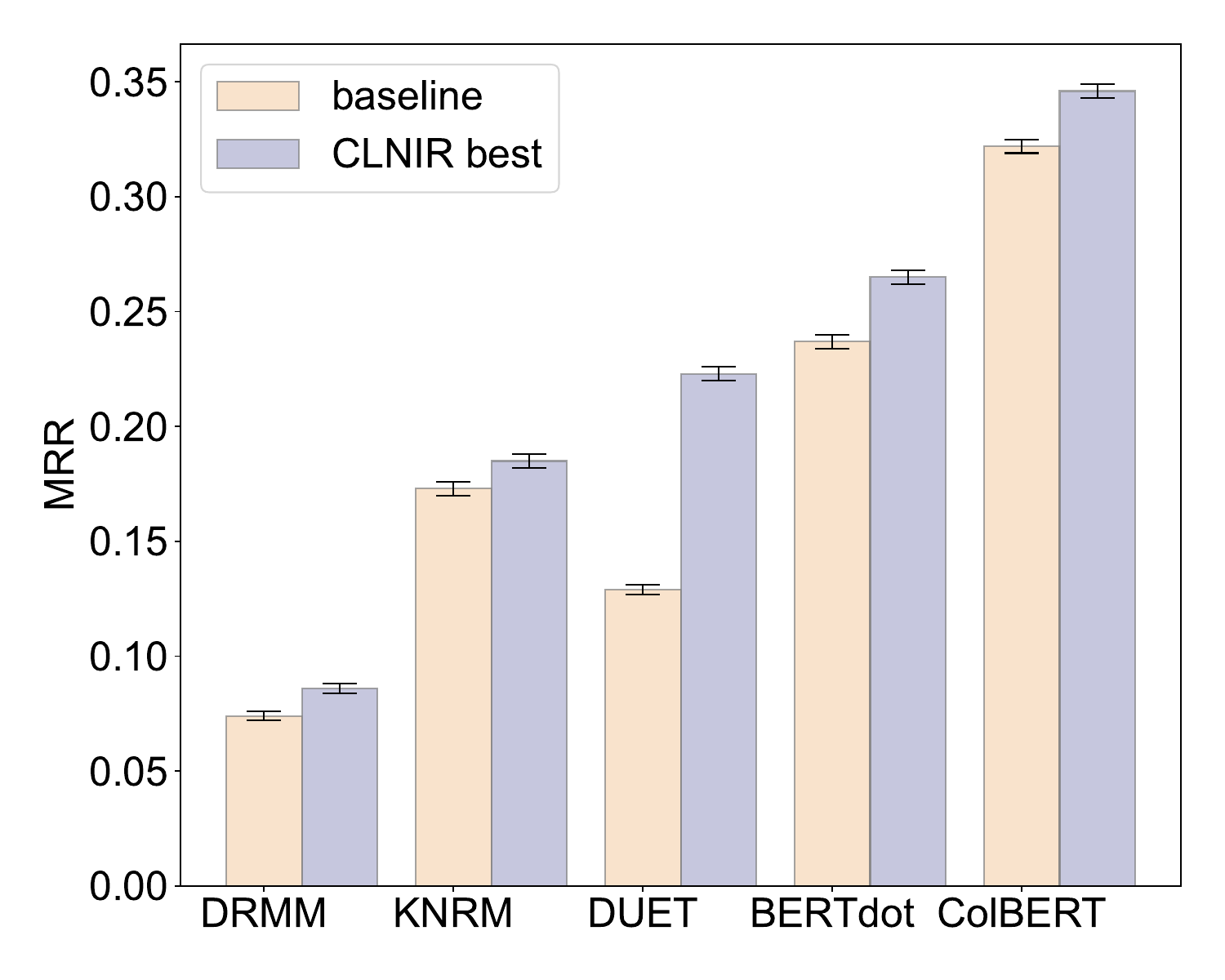}
		\caption{Average final \gls{mrr}}
     	\label{figure.7(a)}
	\end{subfigure}
    \hfil
	\begin{subfigure}{0.32\textwidth}
	   \centering
\includegraphics[width=\textwidth]{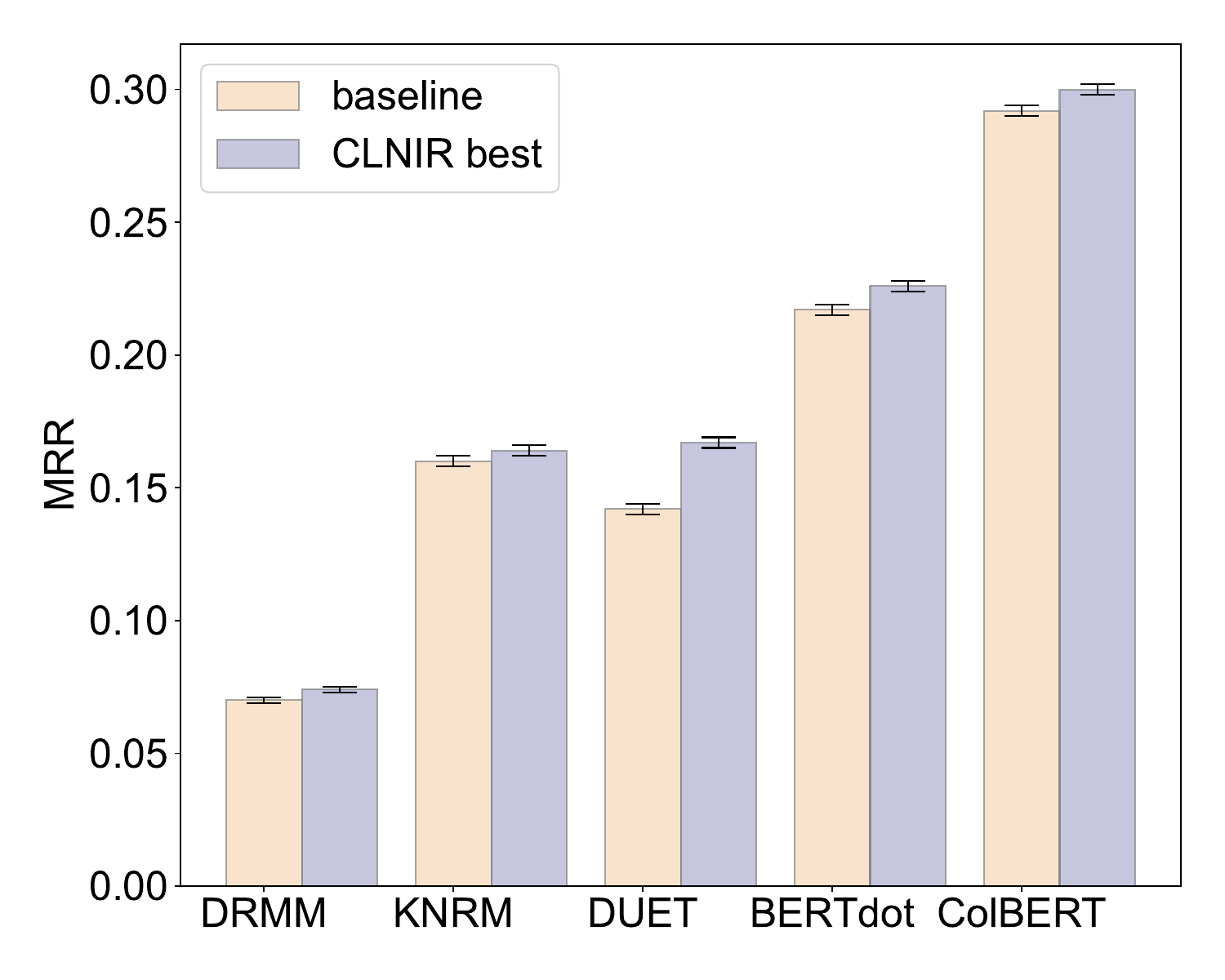}
	   \caption{Forward transfer ability} 
          \label{figure.7(b)}
	\end{subfigure}
        \hfill
  	\begin{subfigure}{0.32\textwidth}
		\centering
		\includegraphics[width=\textwidth]{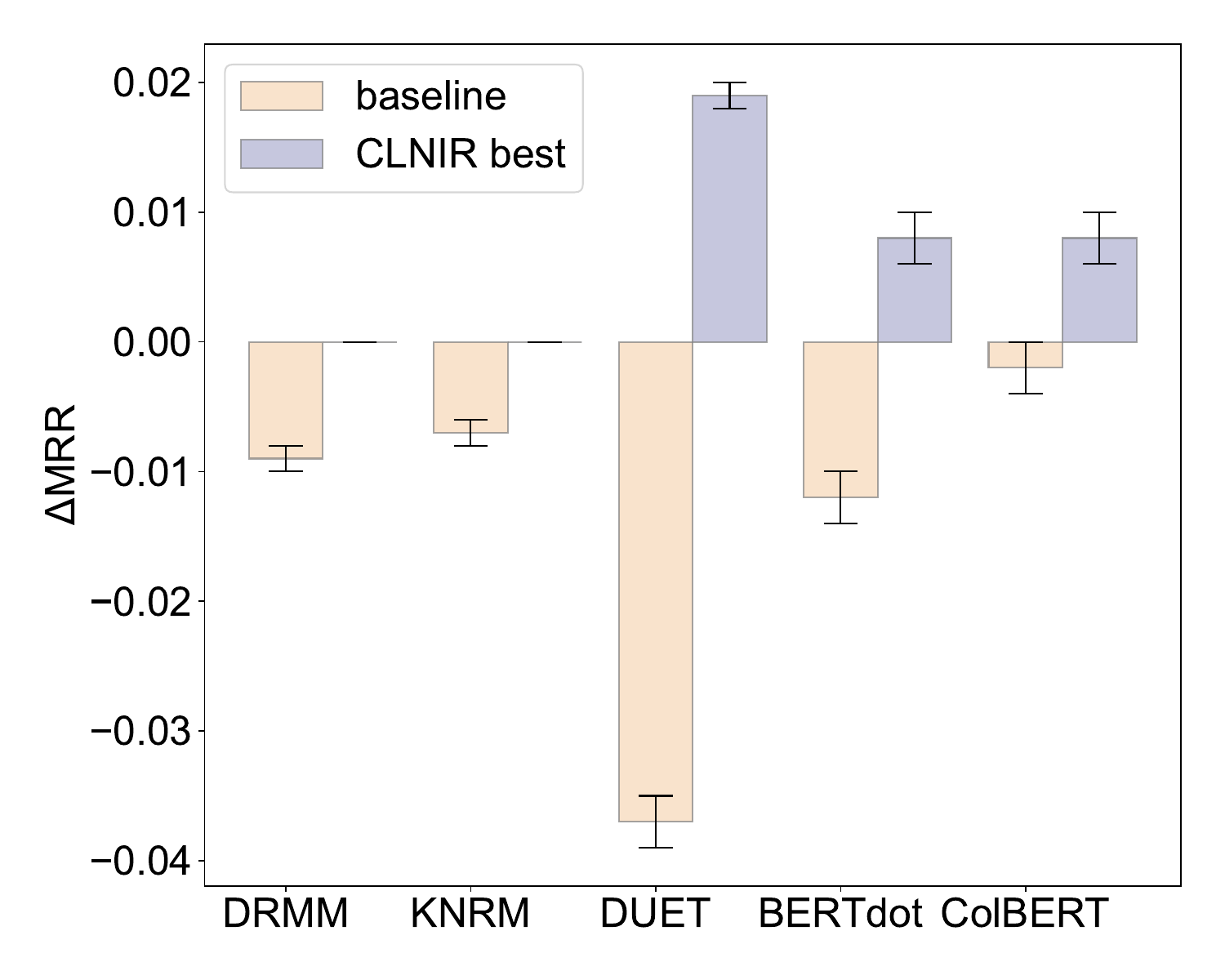} 
		\caption{Backward transfer ability} 
            \label{figure.7(c)}
		\end{subfigure}
      \caption{Performance comparison between baseline and \gls{clnir} best.}  
    \label{figure.7}
	\end{figure}

To provide further insight into the effectiveness of learning strategies, the performance scores of three metrics on the five \gls{nir} models are depicted in Figure~\ref{figure.7}. For ease of comparison, the combination of \gls{nir} models and their best corresponding learning strategies will be referred to as ``\gls{clnir} best''.

\subsection{Relationship between data volume/topic shift and continual \gls{nir}}

This discusses the relationship between the performance of continual information retrieval on the one hand and topic shift and data volume on the other.

\subsubsection{Topic shift}

\textbf{Aim and methodology of the experiment.} The experiment aims to explore the impact of topic shift distances on the performance of \gls{nir} tasks in continual learning. The methodology is as follows:
\begin{itemize}
\item  Topic t1 (``furnishing'') was selected as the starting point because it has the smallest mean standard deviation among the other topics, resulting in the most uniform topic distances according to Figure~\ref{figure.3(b)}.
\item  The remaining topics (t0, t2--t5) were learned in parallel as the second topics. Therefore, five sub-experiments were conducted, each with two topics.
\item  The \gls{nir} models used for the experiment were \gls{knrm} and \gls{colbert}, as their baseline methods performed the best among embedding-based and pretraining-based models, respectively.
\item  The continual learning strategy used was the \gls{si} strategy, as it achieved the most non-negative \gls{fwt} scores in Table~\ref{table.3}.
\end{itemize}

\begin{figure}[htbp]
\label{figure.8}
\begin{subfigure}{0.48\textwidth}
    \centering
    \includegraphics[scale=0.25]{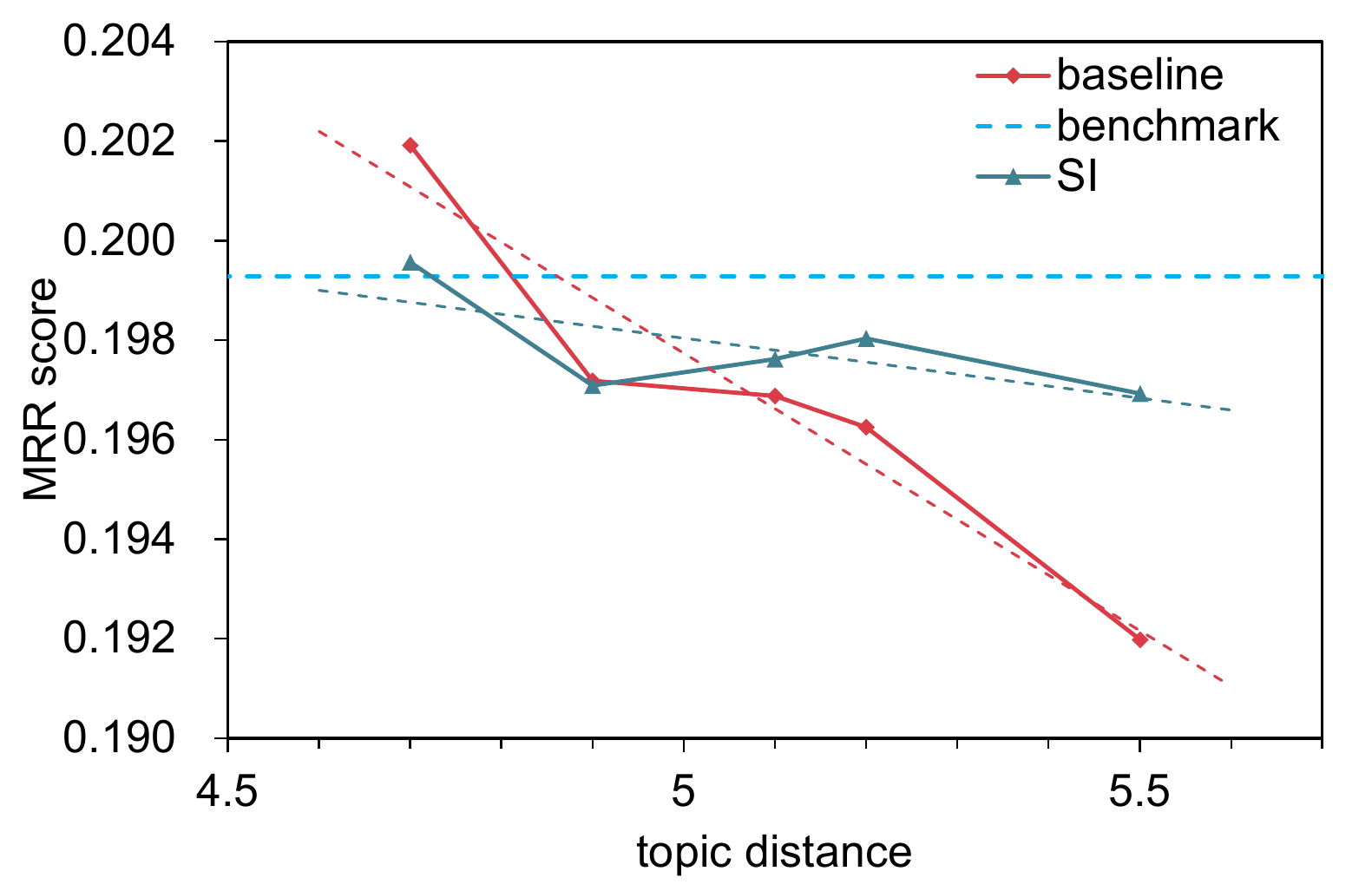} 
    \caption{Topic shift phenomenon of \gls{knrm} model}
    \label{figure.8(a)}
\end{subfigure}
    \hfill
\begin{subfigure}{0.48\textwidth}
    \centering
    \includegraphics[scale=0.25] {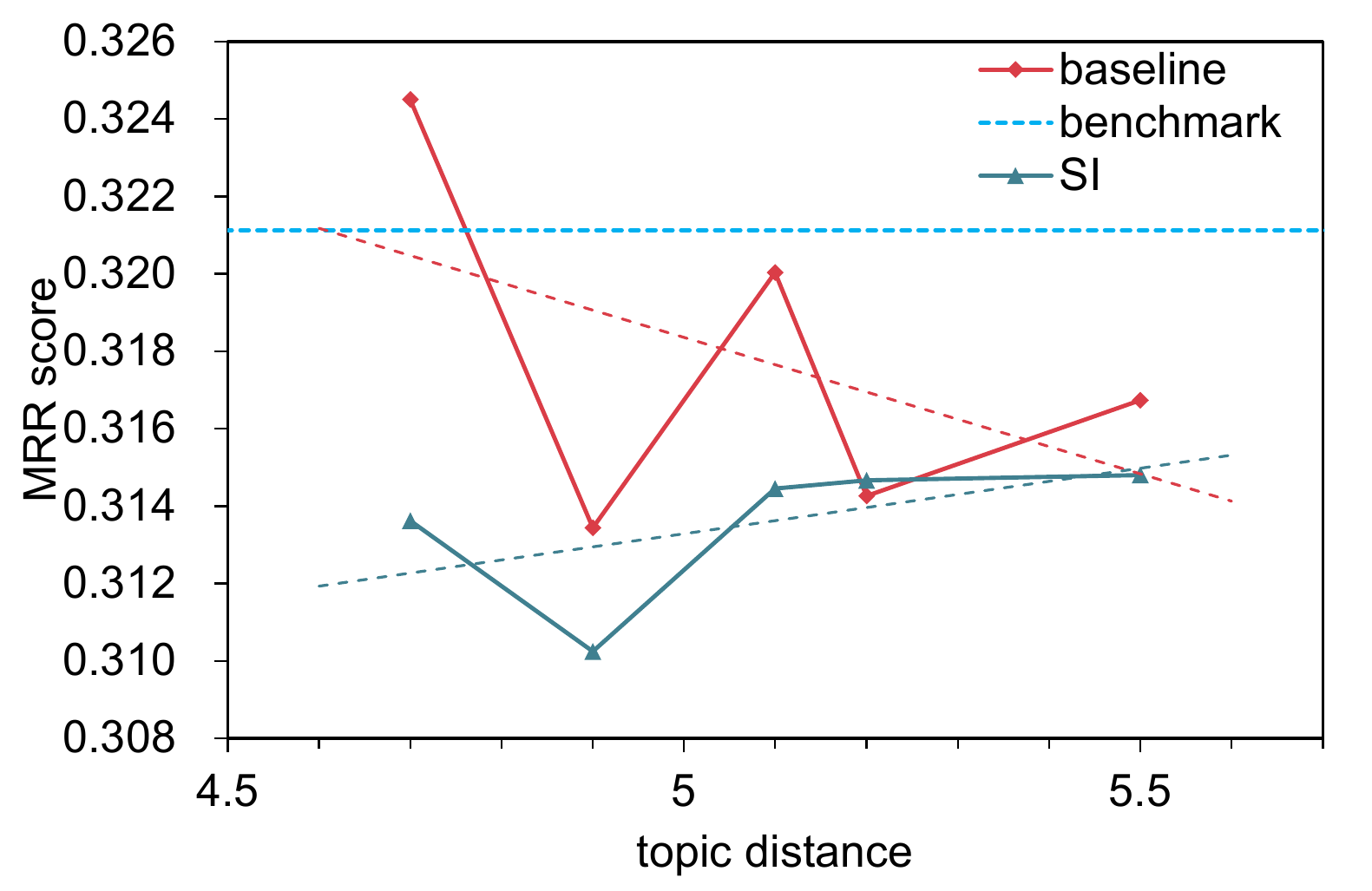}
    \caption{Topic shift phenomenon of \gls{colbert} model} 
    \label{figure.8(b)}
\end{subfigure}
\caption{Impact of topic shift on continual learning performance.
  The ``benchmark'' refers to the results obtained prior to learning the second task; the ``baseline'' represents the results of \gls{nir} models without implementing any learning strategies; and ``SI'' refers to the results of models that have been combined with  \gls{si} strategy.
  } 
\end{figure}

\textbf{Analysing the topic shift in the \gls{knrm} group.} According to Figure~\ref{figure.8(a)}, the baseline \gls{knrm} model without a learning strategy shows that as the topic distance increases, the \gls{mrr} score decreases. The Pearson correlation coefficient between topic distances and \gls{mrr} scores is -0.953. However, when combined with the \gls{si} strategy, the decrease in performance is less severe because the performance curve of the \gls{si} strategy has a slower decline rate compared to the baseline. This implies that the \gls{si} strategy can effectively mitigate the issue of catastrophic forgetting while learning a new task, especially when the topic distance is greater than 4.9. In such cases, the \gls{mrr} scores generated with the \gls{si} strategy are both more stable and higher than those of the baseline methods.

\textbf{Analysing the topic shift in the \gls{colbert} group.} Figure~\ref{figure.8(b)} demonstrates that  \gls{colbert} produced different results compared to the \gls{knrm} group. The \gls{mrr} performance curve for the baseline  \gls{colbert} displays fluctuations as the topic distance increases, resulting in a general decline in \gls{mrr} scores. On the other hand, the \gls{mrr} curve for the  \gls{colbert} combined with \gls{si} shows a narrower range of fluctuation, however, the \gls{mrr} scores are lower than those of the baseline \gls{colbert}. This suggests that \gls{colbert} does not consistently show a decrease in \gls{mrr} scores with increasing topic distance. Furthermore, when combined with the \gls{si} strategy, the majority of the \gls{mrr} scores are lower than the baseline.

\subsubsection{Data volume}

\textbf{Aim and methodology of the experiment.}
This experiment aims to investigate the relationship between data volume and performance. The methodology for conducting this experiment is described below:
\begin{itemize}
\item \gls{knrm} and \gls{colbert} were chosen as the tested \gls{nir} models and the \gls{si} strategy was employed. The reasons for these choices are as previously stated.
\item The initial task was selected to be ``tourism'' (t4) and the subsequent task was ``food'' (t2), as these two topics were deemed to have the greatest disparity among all topic pairs.  
\item The influence of varying training sample sizes on model performance was assessed by setting the sample size for the second task in a flexible manner. Considering that the sample size for t2 is approximately five times larger than that of t4, the sample size for the second task was established to range from 0.1 to 1 times, incrementing by 0.1 at each interval, and then from 1 to 5 times with an interval of 1.
\end{itemize}

\begin{figure}[ht]
        \label{figure.9}
		\begin{subfigure}{0.5\textwidth}
			\centering
			\includegraphics[scale=0.25]{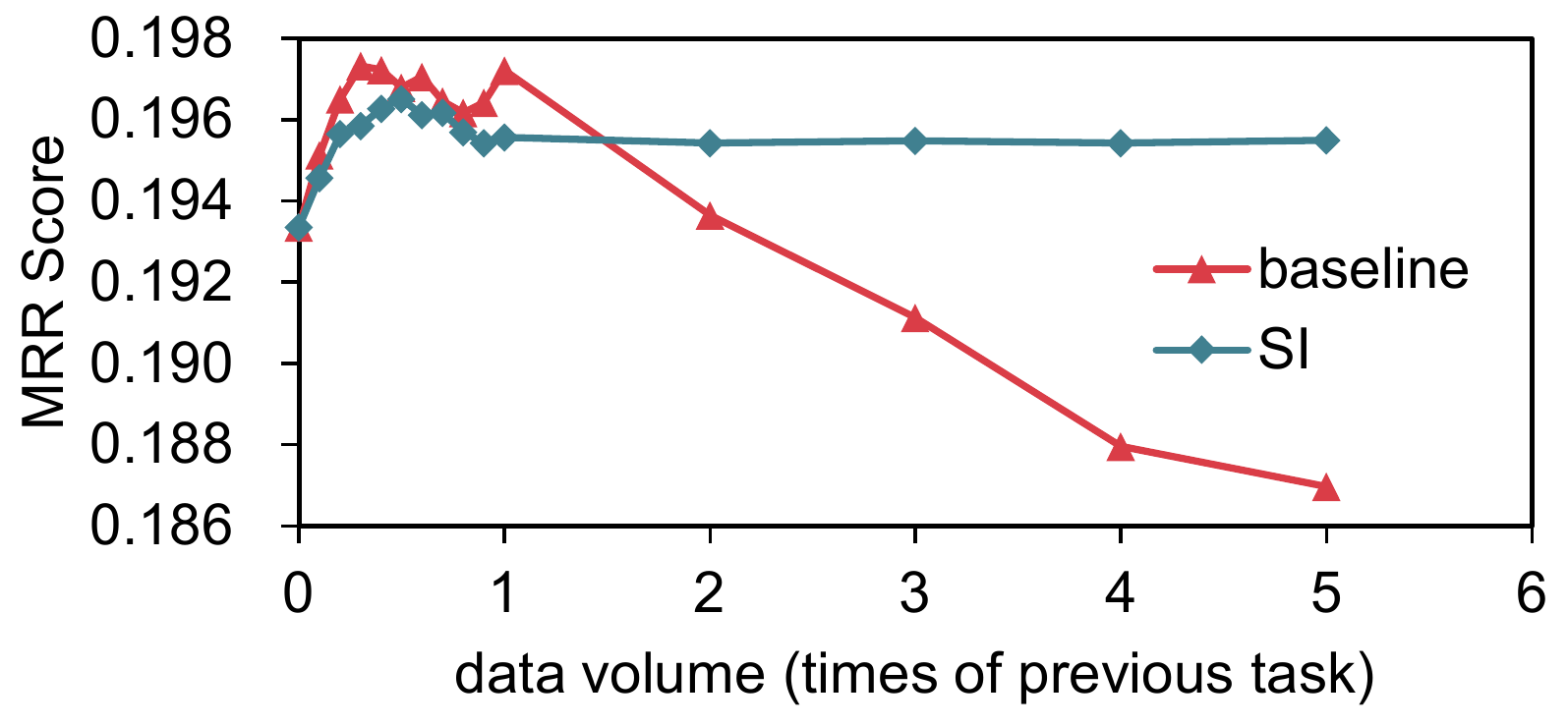} 
			\caption{Results of the \gls{knrm} group}
   	\label{figure.9(a)}
		\end{subfigure}
        \hfill
		\begin{subfigure}{0.5\textwidth}
			\centering
			\includegraphics[scale=0.25]{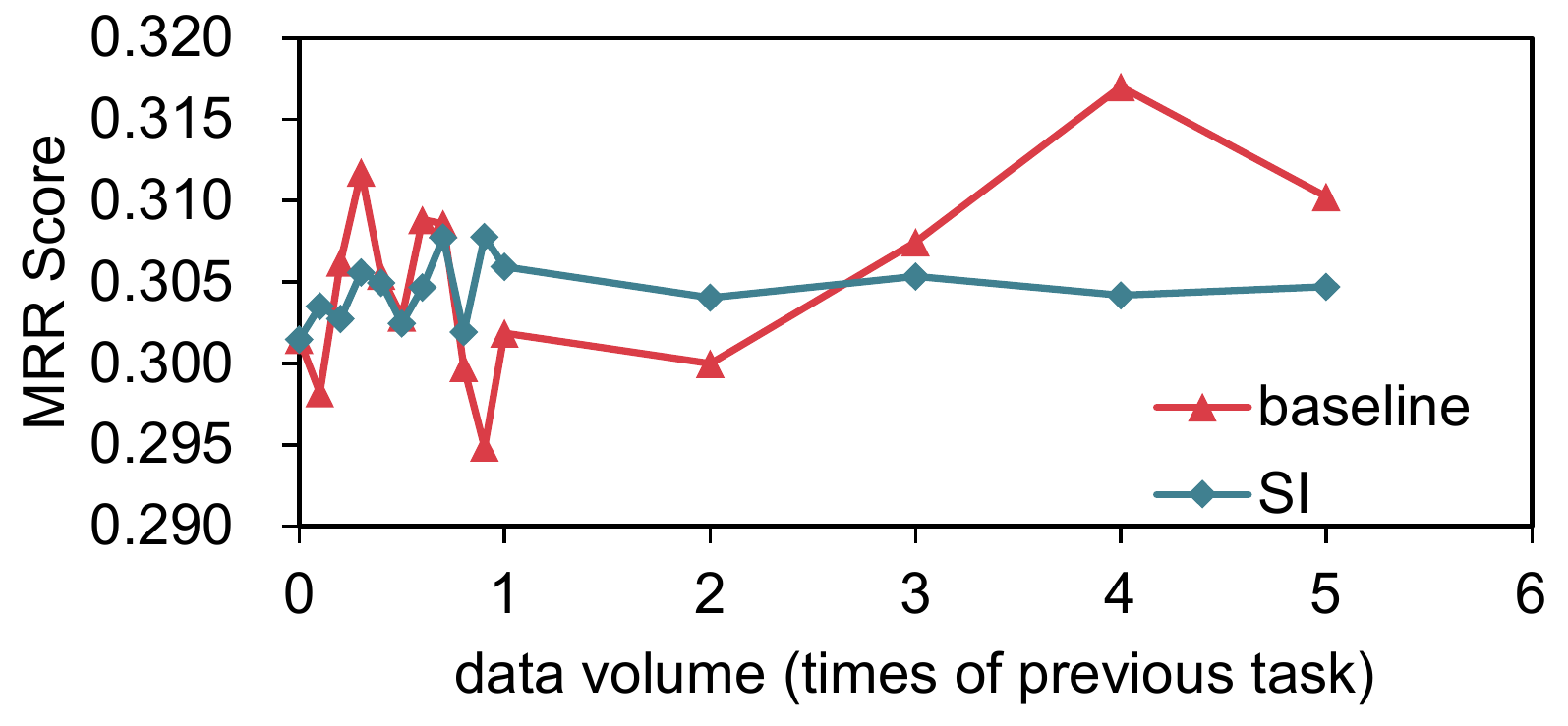}
			\caption{Results of the \gls{colbert} group.} 
           \label{figure.9(b)}
		\end{subfigure}
  		\caption{Impact of data volume on continual learning performance}
	\end{figure}

\textbf{\textbf{Analysing the data volume in the \gls{knrm} group.}} The relationship between data volume and the performance of \gls{knrm} methods is demonstrated in Figure~\ref{figure.9(a)}.
It has been observed that in the baseline \gls{knrm} method, there exists a negative correlation between data volume and continual learning performance, where the data volume of the new task is one time larger. This correlation is measured using the Pearson correlation coefficient of -0.987. On the other hand, the \gls{si}-combined \gls{knrm} method exhibits more stable results, indicating that the \gls{si} strategy effectively reduces the impact of data volume. It should be noted that both \gls{knrm} methods exhibit an upward trend when the sample size of the subsequent task is smaller than the initial task, which suggests that embedding-based models can still benefit from small-scale new tasks even when there is a significant topic distance.

\textbf{\textbf{Analysing the data volume in the \gls{colbert} group.}} The results for the \gls{colbert} group are depicted in Figure~\ref{figure.9(b)}. It is shown that the \gls{si}-combined \gls{colbert} method provides more stable performance compared to the baseline \gls{colbert} method, but it failed to achieve higher scores. Both methods have local high points when the sample size of the subsequent task is one time smaller, which is similar to the findings in the \gls{knrm} group.
However, the baseline \gls{colbert} method demonstrates a weaker correlation with data volume, with a Pearson correlation coefficient of 0.490. This is further evidenced by the fluctuating performance curve.

\subsection{Key findings}
Three experiments were conducted to evaluate the performance of the proposed \gls{clnir} framework and reveal the relationships between continual learning performance, topic shift, and data volume.

\textbf{Findings about the continual learning performance of \gls{clnir}.} The findings of the first experiment showed that pretraining-based \gls{nir} models outperformed embedding-based models in terms of average final performance and forward transfer performance. The forward transfer performance rank order was similar to that of the final average performance. For pretraining-based \gls{nir} models, the \gls{nr} strategy was found to be the most effective, while the best strategy for embedding-based \gls{nir} models varied. However, in some instances, the \gls{gem}-combined \gls{nir} models performed worse than the baseline approaches. These results demonstrate the effectiveness of the proposed \gls{clnir} framework in mitigating the catastrophic forgetting phenomenon and improving continual learning performance in \gls{nir} tasks.

 \textbf{Findings about topic shift and data augmentation.} In the second and third experiments, it was observed that both topic shift distance and data augmentation have a negative correlation with continual learning performance when using the embedding-based baseline method. As the topic shift distance or data volume increases, the continual learning performance tends to decrease. However, the use of a learning strategy such as \gls{si} can mitigate this trend. This pattern is also evident in the pretraining-based baseline methods, where \gls{mrr} scores fluctuate with the increment of topic shift distance or data volume, but models combined with \gls{si} are consistently more stable than those baseline models. While learning strategies can address the issue of catastrophic forgetting caused by topic shifts or data augmentation, they may also diminish the performance improvements caused by a new task.

\section{Conclusion}\label{section6}
This paper presents a task definition and novel dataset for continual \glsdesc{nir} tasks and introduces a framework called \gls{clnir} for addressing the lack of continual learning ability in current \gls{nir} models. The \gls{clnir} includes several typical \gls{nir} models and learning strategies. Experiments were conducted to evaluate the performance of different pair combinations of \gls{nir} models and learning strategies in the framework. The results show that the best learning strategies for different \gls{nir} models vary, but \gls{nir} models combined with a suitable learning strategy can always outperform the baseline methods where no learning strategy is used. Additionally, the impact of topic shift and data augmentation has also been investigated through experiments. It has been found that learning strategies can mitigate the impact of topic shift and data augmentation, leading to more stable performance. In conclusion, the proposed \gls{clnir} framework can effectively deal with the catastrophic forgetting problem in \gls{nir}. The proposed framework can even improve the performance of old tasks by learning new tasks. 

The primary limitations of this study include the scope of continual learning strategies employed. Specifically, the investigation was restricted to seven strategies based predominantly on regularization and replay mechanisms, excluding other approaches such as optimization-based and architecture-based strategies~\citep{10444954}. Moreover, the six thematic tasks utilized in this research all originate from MS MARCO, a question-answering retrieval dataset. This singular data source may not fully represent the performance capabilities of the framework across diverse new information retrieval tasks, such as recommender systems. Lastly, due to computational constraints, a shared bi-encoder was used for both queries and documents within the experiments. Using other NIR architectures, such as cross-encoder or phrase-level representation~\citep{zhao2024dense}, may lead to subtly different experimental observations.

Future research directions can address the current study’s limitations and expand the utility of the \gls{clnir} framework in several ways:
\begin{itemize}
    \item {\bf Development of novel continual learning strategies.} The continual learning strategies in the \gls{clnir} framework are adapted from other domains. This study analyzed their performance on different \gls{nir} models and concluded that no single strategy outperforms others across all models. Therefore, future efforts could concentrate on crafting innovative strategies custom-designed to tackle the unique challenges of information retrieval.
    \item {\bf Expansion to multimodal information retrieval.} With the advent of large-scale multimodal models~\cite{10386743}, the capabilities for multimodal information retrieval have been significantly enhanced. The current framework, which operates solely on text modality, is poised for expansion into multimodal information retrieval. Future studies can further explore this potential, assessing how the \gls{clnir} framework can incorporate and leverage various modalities such as images, videos, and audio for more comprehensive retrieval tasks.
\end{itemize}

\section*{Data availability}

The data described in this paper was obtained from the public repository available at \href{https://microsoft.github.io/msmarco/}{https://microsoft.github.io/msmarco/}. To reorganize the dataset and reproduce the experiments, please refer to the paper's GitHub repository located at
\href{https://github.com/JingruiHou/CLNIR}{https://github.com/JingruiHou/CLNIR}. 

\section*{Acknowledgments}
Jingrui Hou was supported by funding from CSC (China Scholarship Council, No.202208060371) and Loughborough University.

\end{document}